\documentclass{IEEEtran}
\usepackage[utf8]{inputenc}
\usepackage[english]{babel}
\usepackage{amsmath}
\usepackage{amsthm}
\usepackage{amssymb}

\usepackage[final]{changes}

\makeatletter
\def\thickhline{%
  \noalign{\ifnum0=`}\fi\hrule \@height \thickarrayrulewidth \futurelet
   \reserved@a\@xthickhline}
\def\@xthickhline{\ifx\reserved@a\thickhline
               \vskip\doublerulesep
               \vskip-\thickarrayrulewidth
             \fi
      \ifnum0=`{\fi}}
\makeatother

\newlength{\thickarrayrulewidth}
\usepackage{mathtools}

\DeclarePairedDelimiter\floor{\lfloor}{\rfloor}

\usepackage{graphicx}
\usepackage{caption,subcaption}

\usepackage{subfiles}

\usepackage[nottoc]{tocbibind}

\usepackage{tikz}
\usepackage{pgfplots}

\usepackage{color}
\newcommand*{\sarit}[1]{\textcolor{red}{ \textbf{Sarit's comment:} #1}}

\newcommand*{\Sindri}[1]{\textcolor{olive}{ \textbf{Sindri's comment:} #1}}
\newcommand*{\SM}[1]{\textcolor{blue}{#1}}

\usepackage[ruled]{algorithm2e}
\SetKw{KwInput}{Input}
\SetKw{KwReturn}{return}

\newtheorem{theorem}{Theorem}
\newtheorem{lemma}{Lemma}
\newtheorem{definition}{Definition}
\newtheorem{assumption}{Assumption}

\newtheorem{example}{Example}
\newtheorem{remark}{Remark}
\renewcommand{\vec}[1]{{\bf{#1}}}

\title{Compressed Gradient Methods with Hessian-Aided Error Compensation }
\author{Sarit Khirirat, Sindri Magnússon, and Mikael Johansson \thanks{This work was partially supported by the Wallenberg Artificial Intelligence, Autonomous Systems and Software Program (WASP) funded by Knut and Alice Wallenberg Foundation.  S. Khirirat, S. Magnússon and M. Johansson are with the Department of Automatic Control, School of Electrical Engineering and ACCESS Linnaeus Center, Royal Institute of Technology (KTH), Stockholm, Sweden. Emails: \{\textit{sarit@kth.se,sindrim@kth.se, mikaelj@kth.se}\}.} \\ \thanks{A preliminary version of this work is published in the proceedings of IEEE International Conference on Acoustics, Speech, and Signal Processing (ICASSP), 2019 \cite{ourICASSPworks}.}}
\date{}

\begin{document}
 
\maketitle

\begin{abstract}
The emergence of big data has caused a dramatic shift in the operating regime for optimization
algorithms. The performance bottleneck, which used to be computations, is now often communications.  Several gradient compression techniques have been proposed to reduce the communication load at the price of a loss in solution accuracy.  Recently,  it has been shown how compression errors can be compensated for in the  optimization algorithm to improve the solution accuracy. Even though convergence guarantees for  error-compensated algorithms have been established, there is very limited theoretical support for quantifying the observed improvements in solution accuracy. In this paper, we show that Hessian-aided error compensation, unlike other existing schemes, avoids accumulation of compression errors on quadratic problems. We also present strong convergence guarantees of Hessian-based error compensation for stochastic gradient descent. Our numerical experiments highlight the benefits of Hessian-based error compensation, and demonstrate that similar convergence improvements are attained when only a diagonal Hessian approximation is used.

\end{abstract}

\section{Introduction}\label{sec:Intro}

%


 Large-scale and data-intensive problems in machine learning, signal processing, and control are typically solved by parallel/distributed optimization algorithms. 
%
 These algorithms achieve high performance by splitting the computation load between multiple nodes that cooperatively determine the optimal solution. 
  In the process\added{,}  much of the algorithm complexity is shifted from the computation to the coordination.
  This means that the communication can easily become the main bottleneck of the algorithms,  making it expensive to  exchange  full  precision information especially when the decision vectors are large and dense. 
  For example, in training state-of-the-art deep neural network models with millions of parameters such as AlexNet, ResNet and LSTM communication \replaced{can account for}{accounts for} \added{up to} $80\%$ of overall training time, \cite{alistarh2017qsgd,lin2017deep,seide20141}. 
  
%
 %
  


 To reduce the communication overhead in large-scale optimization much recent literature has focused on algorithms that compress the communicated information. 
 Some successful examples of such compression strategies are  \emph{sparsification}\added{, where} some elements of information are set to be zero   \cite{wangni2017gradient,gradientcompression2018}  and \emph{quantization}\added{,} where information is reduced \replaced{to a}{into the} low-precision representation  \cite{alistarh2017qsgd,magnusson2017convergence}.  
%
%
\textcolor{black}{
Algorithms that compress information in this manner have been extensively analyzed for both centralized and decentralized architectures,
\cite{alistarh2017qsgd,gradientcompression2018,magnusson2017convergence,magnusson2019maintaining,khirirat2018distributed,koloskova2019decentralized,doan2018fast,reisizadeh2019robust,zhang2019compressed,reisizadeh2019exact}. 
}
These algorithms are theoretically shown to converge to approximate optimal solutions with \replaced{an}{the} accuracy \replaced{that}{which} is limited by the compression precision.  Even though compression schemes reduce the number of communicated bits in practice, they often lead to significant performance degradation in terms of both solution accuracy and convergence \added{times}, \cite{seide20141,khirirat2018distributed,2019arXiv190109847P,wu2018error}.


To mitigate these negative effects of information compression \replaced{on optimization algorithms}{in optimization methods}, \added{serveral} error compensation strategies have been \replaced{proposed}{suggested in literature}~\cite{seide20141,strom2015scalable,alistarh2018convergence,wu2018error}. 
In essence, error compensation corrects for the accumulation of many consecutive compression errors by keeping a memory of previous errors. Even though \replaced{very}{too} coarse compressors are used, optimization algorithms using error compensation \replaced{often display the same practical performance as}{still can attain almost the same convergence performance} as \added{algorithms} using full-precision information, \cite{seide20141,strom2015scalable}. 
%
%
%
%
Motivated by these encouraging \replaced{experimental observations}{experiments}, several works have \deleted{later} studied different optimization algorithms with  error compensation,  \cite{alistarh2018convergence,ourICASSPworks,wu2018error,Stich2018,2019arXiv190109847P,tang2019doublesqueeze,tang2019texttt}. 
%
%
However, there are not many theoretical \replaced{studies}{supports} which validate why error compensation exhibits better convergence guarantees than direct compression.
%
 %
For instance, Wu  \emph{et. al} \cite{wu2018error} \replaced{derived}{showed} better worst-case bound guarantees of error compensation as the iteration goes on for quadratic optimization. Karimireddy \emph{et. al} \cite{2019arXiv190109847P} showed that \replaced{binary compression may cause optimization algorithms to diverge, already for one-dimensional problems, but that this can be remedied by error compensation.}{optimization algorithms using binary compression \replaced{may}{might} diverge \replaced{already for}{for the minimization problems over simple} one-dimensional \replaced{problems and how this issue can be overcome by error-compensation.}{linear functions and therefore error feedback overcomes this problem. }}
\replaced{However, we}{We} show in this paper (see Remark \ref{remark:Comparison}) that these methods still accumulate errors\added{,} even for quadratic problems. 

%
%
%

The goal of this paper is develop a better theoretical understanding of error-compensation in compressed gradient methods. Our key results quantify the accuracy gains of error-compensation and prove that Hessian-aided error compensation removes \emph{all} accumulated errors on strongly convex quadratic problems. The improvements in solution accuracy are particularly \added{dramatic} on ill-conditioned problems. We also provide strong theoretical guarantees of error compensation in stochastic gradient descent methods distributed across multiple computing nodes. 
	Numerical experiments confirm the superior performance of Hessian-aided error compensation over existing schemes. In addition, the experiments indicate that error compensation with a diagonal Hessian approximation achieves similar performance improvements as using the full Hessian. 

\textbf{Notation and definitions.} We let $\mathbb{N}$ ,$\mathbb{N}_0$, $\mathbb{Z}$, and $\mathbb{R}$ be the set of natural numbers, the set of natural numbers including zero, the set of integers, and the set of real numbers, respectively. The set $\{0,1,\ldots,T\}$ is denoted by $[0,T].$ For $x\in\mathbb{R}^d$, $\|x\|$  and $\|x \|_1$ are the $\ell_2$ norm and the $\ell_1$ norm, respectively, and $\left\lceil x\right\rceil_+ = \max\{0,x\}$. 
 For a symmetric matrix $A\in R^{d\times d}$, we let $\lambda_1(A), \ldots, \lambda_d(A)$ denote the eigenvalues of $A$ in an increasing order (including multiplicities), and its spectral norm is defined by $\|A\|=\max_i \vert \lambda_i(A) \vert$.
A continuously differentiable function $f:\mathbb{R}^d\rightarrow\mathbb{R}$,
is $\mu$-strongly convex if there exists a positive constant $\mu$  such that 
\begin{align}\label{eqn:mustronglyconvex}
f(y)\geq f(x) + \langle \nabla f(x),y-x\rangle +\frac{\mu}{2}\| y-x\|^2,  \quad \forall x,y.
\end{align}
and $L$-smooth if 
\begin{align}\label{eqn:Lsmooth}
|| \nabla f(y) - \nabla f(x)|| \leq  L ||x-y||, \quad \forall x,y\in\mathbb{R}^d.
\end{align}
%
%
%
\section{Motivation and Preliminary Results}

\begin{figure}[t]
    \centering 
\begin{subfigure}{0.24\textwidth}
  \includegraphics[width=\hsize]{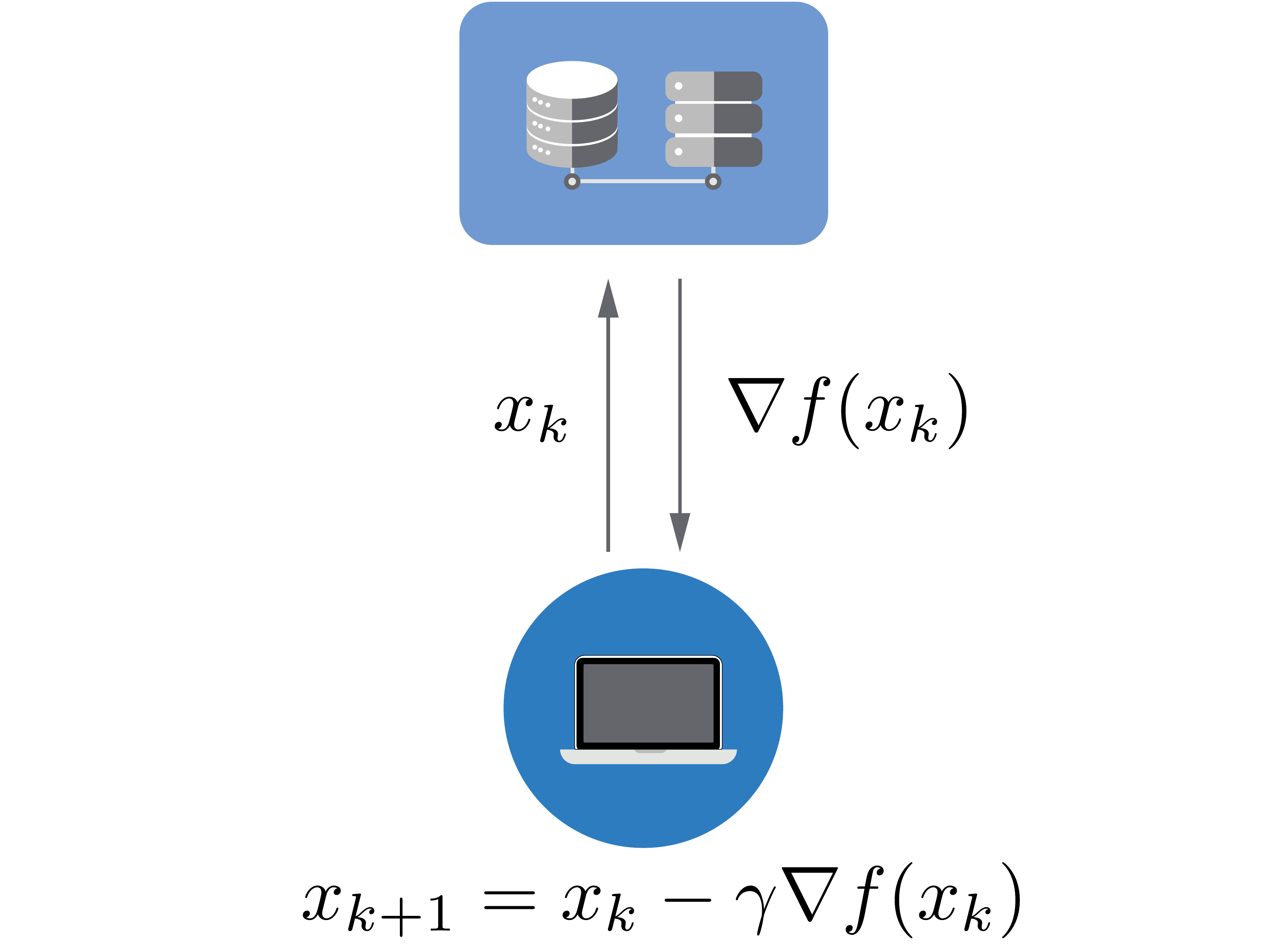}
  \caption{}
  \label{fig:centralized}
\end{subfigure}\hfill 
\begin{subfigure}{0.24\textwidth}
  \includegraphics[width=\hsize]{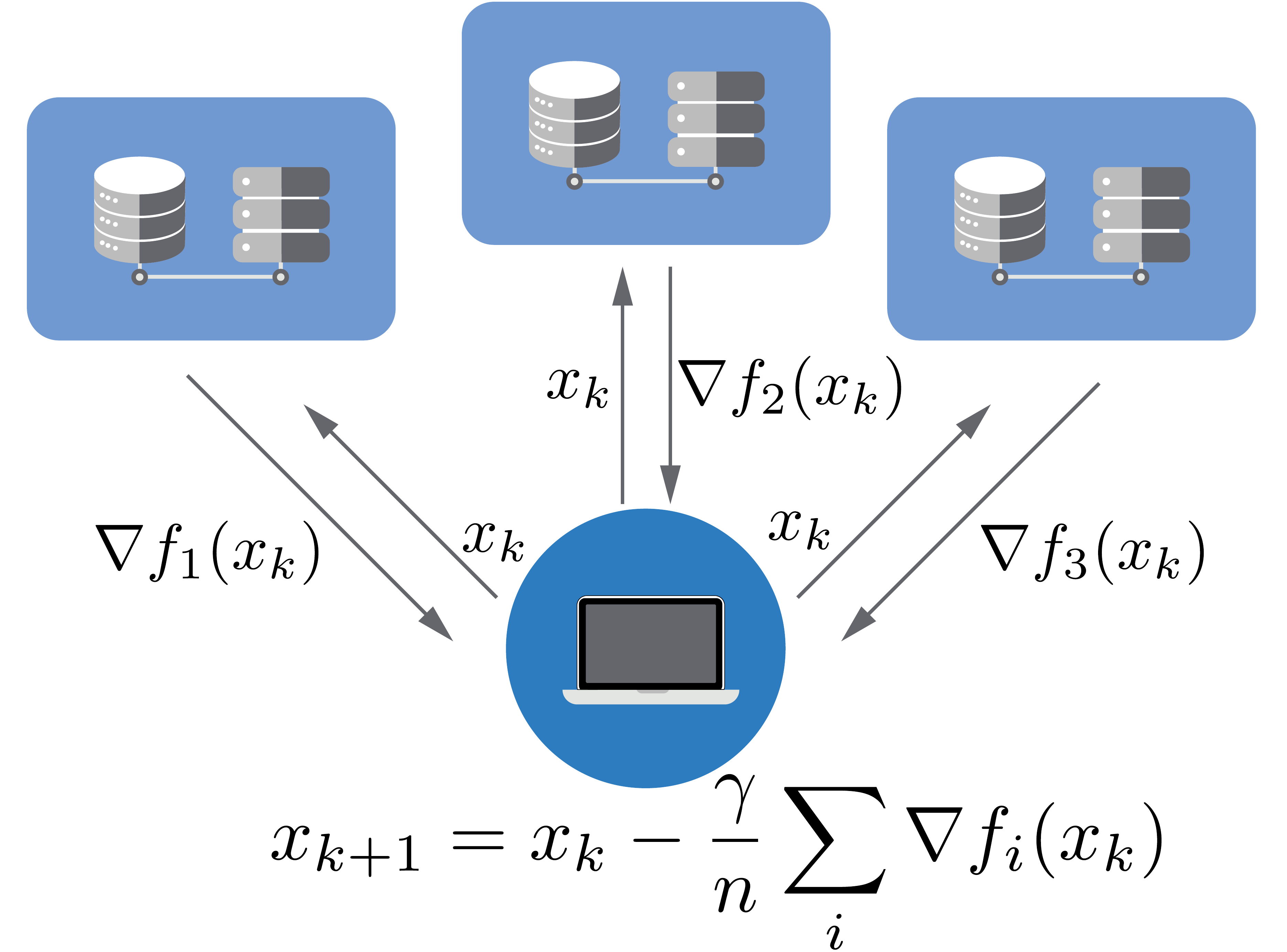}
  \caption{}
  \label{fig:distributed}
\end{subfigure}
\caption{Two common communication architectures for distributed gradient methods: 1) full gradient communication (left)  and 2) partial gradient communication (right).  
}
\label{fig:distributedlearningframework}
\end{figure}

In this section, we motivate our study of error-compensated gradient methods. 
\replaced{We }{ \S~\ref{sec:Mot_Dist} we }give an overview of distributed optimization algorithms  based on communicating gradient information\replaced{ in \S~\ref{sec:Mot_Dist} and describe a}{.}
\deleted{Finally in \S~\ref{sec:Mot_GC} we describe the} general form of gradient \replaced{compressors}{compressions}, covering most \replaced{existing ones,}{ compression schemes} in \replaced{\S~\ref{sec:Mot_GC}.}{the literature. }
Later in \S~\ref{sec:Mot_Low} we illustrate the limits of directly compressing the gradient, motivating the need for the error-compensated gradient methods studied in this paper.

{\color{black}
\subsection{Distributed Optimization}\label{sec:Mot_Dist}
Distributed optimization has enabled  solutions to large problems in many application areas, e.g., smart grids, wireless networks, and statistical learning. Many distributed  algorithms build on gradient methods and can be categorized based on whether they use a) full gradient communication \replaced{or}{and} b) partial gradient communication\replaced{; see}{, as shown in} Figure \ref{fig:distributedlearningframework}. The full gradient \deleted{communication} algorithms solve problems on the form
 \begin{align}\label{eqn:ProblemFullFcn}
\mathop\text{minimize}\limits_{x} \quad f(x), 
\end{align}
 by \replaced{the standard gradient descent iterations}{performing a standard gradient method} 
 \begin{align}\label{eqn:FullGDcentralized}x^{k+1}= x^k-\gamma \nabla f(x^k),\end{align}
  \replaced{communicating}{where} the full gradient $ \nabla f(x^k)$ \replaced{in}{is communicated at} every iteration.
   Such \replaced{a communication pattern}{communication} usually appears in dual decomposition methods where $f(\cdot)$ is a dual function associated with some large-scale primal problems\replaced{;}{,} we illustrate this in subsection~\ref{sec:Mot_Dist1}. 
 The partial gradient \deleted{communication} algorithms \added{are used to} solve separable optimization problems\deleted{ among $n$ nodes} on the form
 \begin{align}\label{eqn:Problem}
\mathop\text{minimize}\limits_{x} \quad f(x) = \frac{1}{n}\sum_{i=1}^n f_i(x),
\end{align}
  by  \replaced{gradient descent}{performing a separable gradient method}
  \begin{align}x^{k+1}=x^k-\frac{\gamma}{n} \sum_{i=1}^n \nabla f_i(x^k) , \label{eqn:fulldistGD}\end{align}
   \replaced{and distributing the gradient evaluation on $n$ nodes, each responsible for evaluating one of the partial gradients $\nabla f_i(x)$}{where each partial gradient $\nabla f_i(x^k)$ is communicated by node $i$}; see \S~\ref{subsec:PartialGradient}). 
  Clearly, \deleted{the} full gradient communication is a special case of \deleted{the} partial gradient communication with $n=1$.  
  However, considering the full gradient communication algorithms separately  will enable us to get stronger results for this scenario. We now review these algorithms separately in \replaced{more detail.}{the next subsections.}

\subsubsection{Full Gradient Communication (Dual Decomposition)} \label{sec:Mot_Dist1}


\replaced{Resource allocation is a class of}{\emph{Resource allocation} is the modern} distributed optimization problem\added{s} where a group of $n$ 
nodes aim to minimize the sum of their local utility functions under \replaced{a set of shared resource constraints}{coupled/shared constraints (\emph{e.g.}, resource limitations)}. In particular, \replaced{the}{all} nodes collaboratively solve 
\begin{equation} \label{EQ:ProblemFullGradient}
\begin{aligned}
& \underset{q_1,\ldots,q_n}{\text{maximize}}
& & \sum_{i=1}^n U_i(q_i)\\
& \text{subject to}
&& q_i\in \mathcal{Q}_i,~~i=1,\ldots,n\\
& &  &h(q_1,q_2,\ldots,q_n) =  0.   
\end{aligned}
\end{equation}
\replaced{Each}{Here, each} node has \replaced{a utility function $U_i(q)$ depending on its own private resource allocation $q_i$, constrained by the set ${\mathcal Q}_i$.}{its private decision variable $q_i$ \replaced{and}{with associated} utility function $U_i(q_i)$ and the constraint set $\mathcal{Q}_i$.} The decision variables are coupled through the \added{total resource} constraint $h(q_1,q_2,\ldots,q_n) =  0$, which captures \added{system-wide} physical or economical limitations. 
 Distributed algorithms for these problems are often based on solving a dual problem on the form~\eqref{eqn:ProblemFullFcn}. Here $x$ is the dual variable (associated to the coupling constraints) which is updated using the dual gradient method~\eqref{eqn:FullGDcentralized}  with $f(x):= \min_{q} L(q,x)$ and
$$L(q,x)= \sum_{i=1}^n U_i(q_i)+x^T h(q_1,\ldots,q_n ),$$
 is the Lagrangian function.  The dual function is convex and the dual gradient (or a dual subgradient) is given by
 $$\nabla f(x)= h(q_1(x),\ldots,q_n(x) ),~~~~q(x)=\underset{q}{\text{argmin}}\, L(q,x).$$
 The dual gradient can often be measured from the effect of the current  decisions ~\cite{low1999optimization,chiang2007layering,palomar2006tutorial,zhao2014design,magnusson2017convergence,magnusson2016distributed}. Therefore, we get a distributed algorithm 
  \begin{align*}
         q_i^{k+1}=&\underset{q}{\text{argmin}} ~U_i(q_i^k) +(x^k)^T  h_i(q_i^k) , ~~~~i=1,\ldots,n \\
        x^{k+1} =& x^k + \gamma \nabla f(x^k)
 \end{align*}
 where the main communication is the transmission of the dual gradient to the users. To communicate the gradient it must first be compressed into the finite number of bits. Our results  demonstrate naive gradient compression can be improved by an error correction step, leading to accuracy improvements.

\subsubsection{Partial Gradient Communication} \label{subsec:PartialGradient}




\deleted{Another setting of distributed optimization is where $n$ nodes wish to solve the separable problem \eqref{eqn:Problem}. }
%
%
%

Problems on the form of (\ref{eqn:Problem}) appear, \emph{e.g.}, in machine learning and signal processing. 
%
%
One important example is 
\emph{empirical risk minimization} (ERM)  where labelled data is split among \deleted{the }$n$ nodes which collaborate to find the optimal estimate.  
In particular,  if each node $i\in [1,n]$ has access to its local data with feature vectors $\vec{z}_i=(z_i^1,\ldots,z_i^m)$ and labels $\vec{y}_i=(y_i^1,\ldots,y_i^m)$ with $z_i^j\in \mathbb{R}^d$ and $ y_i^j\in \mathbb{R}$, then the local objective functions \replaced{are defined as}{ have the form}
\begin{align}\label{eqn:fi_ERM}
f_i(x){=}\frac{1}{m}\sum_{j=1}^m \ell(x; z_i^j,y_i^j)+ \frac{\lambda}{2}||x||^2, ~ \text{for } i=1,2,\ldots,n
\end{align}
where $\ell(\cdot)$ \replaced{is}{are} some loss function\deleted{s} and $\lambda>0$ is \replaced{a}{the} regularization parameter.  The ERM formulation covers many important machine learning problems: least-squares regression if $\ell(x; z,y)=  (1/2)(y- z^T x)^2$; the logistic regression problem \replaced{when}{if} $\ell(x; z,y)=\log(1+\exp(-y\cdot z^T x)),$; 
 and support vector machines (SVM\deleted{s})  if  $\ell(x; z,y)= \left\lceil1-y\cdot z^T x \right\rceil_+.$


When the data set on each node is large,\deleted{then} the above optimization problem is typically solved \replaced{using}{with} stochastic gradient decent (SGD).
%
%
%
%
%
%
 In each SGD iteration, the master node broadcasts a decision variable $x^k$, while each worker node $i$ computes a stochastic gradient 
$ g_i(x^k;\xi_i^k)$  based on a random subset of its local data ${\mathcal D}_i$.
Then, the master performs the update
\begin{align}\label{eqn:minibatchSGD}
x^{k+1}=x^k-\gamma \frac{1}{n}\sum_{i=1}^n g_i(x^k;\xi_i^k).
\end{align}
We assume that the stochastic gradient preserves the unbiasedness and bounded variance assumptions, i.e. 
\begin{align}\label{eqn:unbiasednessSGD}
&\mathbb{E}_{\xi_i}  g_i(x;\xi_i) = \nabla f_i(x), \quad \text{and} \\
&\mathbb{E}_{\xi_i} \|  g_i(x;\xi_i) - \nabla f_i(x) \|^2 \leq \sigma^2, \quad \forall x\in\mathbb{R}^d. \label{eqn:boundedvarianceSGD}
\end{align}
%
%
%
%
%
To save communication\deleted{s} bandwidth, worker nodes need to compress stochastic gradients into low-resolution  representation\added{s}. 
 Our results illustrate how the low-resolution gradients can achieve  high accuracy solutions by  error-compensation. 

}

\subsection{Gradient Compression} \label{sec:Mot_GC}


\replaced{We consider the following class of gradient compressors.}{To study \replaced{the effect}{effects} of error compensation mechanisms on compressed gradient methods, we consider general compressors having the following property.}

\begin{definition}\label{def:DBEC} 
The operator $Q:\mathbb{R}^d\rightarrow\mathbb{R}^d$ is an $\epsilon$-\emph{compressor} if there exists a positive constant $\epsilon$ such that
\begin{align*}
\| Q(v) - v \| \leq \epsilon, \quad \forall v\in\mathbb{R}^d.
\end{align*}
\end{definition}


\deleted{From} Definition \ref{def:DBEC} \replaced{only requires}{we require only the} bounded magnitude of \added{the} compression error\added{s}\replaced{. A small value of $\epsilon$ corresponds to high accuracy. At the extreme when $\epsilon=0$, we have $Q(v)=v$.}{, which  implies high compression accuracy if $\epsilon$ has low value (e.g. $Q(v)=v$ when $\epsilon=0$)} \replaced{An $\epsilon$-compressor does not need to be unbiased (in constrast to those considered in~ \cite{alistarh2017qsgd,khirirat2018distributed})}{The compressors according to Definition \ref{def:DBEC} are not require to be unbiased unlike  \cite{alistarh2017qsgd,khirirat2018distributed},} and \replaced{is}{are} allowed to have \added{a} quantization error arbitrarily larger than magnitude of the original vector \replaced{(in constrast to~\cite[Definition 2.1]{Stich2018} and \cite[Assumption A]{2019arXiv190109847P})}{unlike \cite[Definition 2.1]{Stich2018} and \cite[Assumption A]{2019arXiv190109847P}}. Definition \ref{def:DBEC} covers \added{most} popular compressors in machine learning and signal processing appplications, which substantiates the generality of our results later in the paper. 
One common example is \replaced{the}{a} rounding quantizer, where  each element of a full-precision vector $v_i\in\mathbb{R}$ is rounded to the closet point in a grid with  resolution level $\Delta>0$ \deleted{according to} 
\begin{align}\label{eqn:RoundingQuant}
[Q_{dr}(v)]_i &= {\rm sign}(v_i) \cdot \Delta\cdot \floor*{ \frac{|v_i|}{\Delta} +\frac{1}{2}}.
\end{align}
%
%
%
%
%
%
This rounding quantizer  is a  $\epsilon$-compressor with $\epsilon = d\cdot\Delta^2/4$, \cite{rabbat2005quantized,zhu2016quantized,li2017training,de2018high}. 
In addition, if \replaced{gradients are bounded, }{we assume bounded gradient assumptions and apply the compressor on gradients, then }the sign compressor \cite{seide20141}, the $K$-greedy quantizer \cite{gradientcompression2018} and the dynamic gradient quantizer \cite{alistarh2017qsgd,gradientcompression2018} \replaced{are all}{all are} $\epsilon$- compressors.




\section{The Limits of Direct Gradient Compression} 
\label{sec:Mot_Low}


To reduce communication overhead in distributed optimization, it is most straightforward to compress the gradients directly.  The goal of this section is to illustrate the limits of this approach, which motivates our gradient correction compression algorithms in the next section.

\subsection{Full Gradient Communication and Quadratic Case}


\replaced{A major drawback with direct gradient compression is that it leads to error accumulation.}{One major drawback of directly compressing the gradients is that it leads to accumulation of residual errors} To illustrate why this happens \deleted{in an intuitive manner} we start by considering \replaced{convex quadratic objectives}{the minimization problems over the quadratic objective} 
\begin{align} \label{eq:quadOBJ}
     f(x)=\frac{1}{2}x^T Hx + b^Tx. 
\end{align}
 \replaced{Gradient descent using compressed gradients reduces to}{Now gradient descent where the gradient information is compressed reduces to the following iterations. }
\begin{align}\label{eqn:CGD}
x^{k+1} = x^k - \gamma Q\left(  \nabla f(x^k) \right),
\end{align}
\replaced{which can  be equivalently expressed as}{or equivalently in this case}
\begin{align} \label{eq:xupdate}
   x^{k+1} =& \overbrace{(I-\gamma H)}^{:=A_{\gamma}}x^k -\gamma b+\gamma \left( \nabla f(x^k)-Q\left(\nabla f(x^k) \right) \right).
\end{align}
\replaced{Hence, }{which yields} 
\begin{equation}
\begin{aligned} \label{eq:FGC-QD}
x^{k+1}-x^{\star} &= A_{\gamma}^{k+1}(x^0-x^{\star}) \\
	                       &\hspace{0.4cm}+\gamma \sum_{j=0}^k A_{\gamma}^{k-j} \left( \nabla f(x^j)-Q\left(\nabla f(x^j) \right) \right).
\end{aligned}
\end{equation}
\replaced{where}{Here} $x^{\star}$ is the optimal solution and the equality follows from the fact that $H x^\star + b=0$. The final term of Equation~\eqref{eq:FGC-QD} \replaced{describes how}{is where} the compression errors from every iteration accumulate. We show how error compensation helps \added{to} remove this accumulation  in Section~\ref{sec:ECGC}. Even though the error accumulates, the compression error will remain bounded if the matrix $A_{\gamma}$ is stable (which can be achieved by a \added{sufficiently} small step-size), as illustrated in the following theorem. 
%
 %
 %

\begin{theorem} \label{thm:CGDQuadratic}
Consider the \deleted{quadratic} optimization problem over the objective function~\eqref{eq:quadOBJ} where $H$ is positive definite and let $\mu$ and $L$ be the smallest and largest eigenvalues of $H$, respectively. Then, the iterates $\{x^k\}_{k\in\mathbb{N}}$ generated by \eqref{eqn:CGD} satisfy
\begin{align*}
\| x^k - x^\star \| &\leq \rho^k\| x^0 - x^\star \| + \frac{1}{\mu}\epsilon, 
\end{align*}
where
\begin{align*}
\rho &= \left\{ \begin{array}{ll} 1-1/\kappa & \text{if } \quad \gamma = 1/L \\   1-{2}/{(\kappa+1)} & \text{if }\quad \gamma = 2/(\mu+L)   \end{array}\right.  ,
\end{align*}
and $\kappa = L/\mu$ is the condition number of $H$.
\end{theorem}
\begin{proof}
See Appendix \ref{app:thm:CGDQuadratic}.
\end{proof}
%
%
Theorem \ref{thm:CGDQuadratic} shows that  \added{the iterates of the} compressed gradient descent in Equation~\eqref{eqn:CGD} converge\deleted{s} linearly \replaced{to}{toward solution} with residual error $\epsilon/\mu$\replaced{}{ where $\epsilon$ is the quantization error and $\mu$ is the smallest eigenvalue of $H$ (or the strong convexity parameter of the  quadratic objective function)}. \replaced{The theorem recovers}{In addition, we recover} the results of classical gradient descent when \deleted{we let} $\epsilon=0$\replaced{.}{ in Theorem \ref{thm:CGDQuadratic}.}


We show in Section~\ref{subsec:Limits} that this upper bound is tight.  With our error-compensated method as presented in Section~\ref{sec:ECGC} we can achieve arbitrarily high solution accuracy even for fixed $\epsilon>0$ and $\mu>0$. 
 First we illustrate how to extend these results to include partial gradient communication, stochastic, and non-convex optimization problems as we show next.

\subsection{Partial Gradient Communication} \label{subsec:CpartialGradient}
We now study  direct gradient compression \replaced{in the}{for} partial gradient communication architecture. \replaced{We focus on}{In particular, we consider the} the distributed compressed stochastic gradient descent algorithm (D-CSGD)
\begin{align}\label{eqn:QDGD}
x^{k+1} = x^k - \gamma \frac{1}{n}\sum_{i=1}^n Q( g_i(x^k;\xi_i^k) ),
\end{align}
where each $g_i(x;\xi_i)$ is \replaced{a}{the} partial stochastic gradient sent by worker node $i$ to the central node.
%
%
\textcolor{black}{In the deterministic case, we have the following  \deleted{rate} result\deleted{s} \deleted{(which are} analogous to Theorem~\ref{thm:CGDQuadratic}.}
{\color{black}
\begin{theorem} \label{thm:3}
 Consider the optimization problem \eqref{eqn:Problem} where $f_i(\cdot)$ are $L$-smooth and $f(\cdot)$ is $\mu$-strongly convex. Suppose that $g_i(x^k;\xi_i^k)=\nabla f_i(x^k)$. If $Q(\cdot)$ is the $\epsilon$-compressor and $\gamma = 2/(\mu+L)$ then 
    \begin{align*}
        \|x^k - x^\star \| \leq \rho^k \|x^0 - x^\star \| + \frac{1}{\mu}\epsilon.
    \end{align*}
    where $\rho = 1 - 2/(\kappa+1)$ and $\kappa=L/\mu$. 
\end{theorem}
\begin{proof}
  See Appendix~\ref{App:Pr3}.
\end{proof}
}
 More generally, we have the following result.
%
%
%
%
%
\begin{theorem}\label{thm:distCSGD_convex} 
Consider an optimization problem \eqref{eqn:Problem} where each $f_i(\cdot)$ is $L$-smooth, and the iterates $\{x^k\}_{k\in\mathbb{N}}$ generated by \eqref{eqn:QDGD} under the assumption that the underlying partial stochastic gradients $g_i(x^k, \xi_i^k)$ satisfies the unbiased and bounded variance assumptions in Equation~\eqref{eqn:unbiasednessSGD} and \eqref{eqn:boundedvarianceSGD}. Assume that $Q(\cdot)$ is  the $\epsilon$-compressor and $\gamma < 1/(3L)$. 

\begin{itemize}
\item  [a)] (non-convex problems) Then, 
\begin{equation}\label{eqn:distCSGD_NC}
\begin{aligned}
 \mathop{\min}\limits_{l\in[0,k]} \mathbf{E}\| \nabla f(x^l)\|^2 
&\leq  \frac{1}{k+1}\frac{2}{\gamma}\frac{1}{1-3L\gamma}\left( f(x^0) - f(x^{\star}) \right) \\
&\hspace{0.4cm}+ \frac{3L}{1-3 L\gamma}\gamma \sigma^2 +\frac{1+3L\gamma}{1-3L\gamma}\epsilon^2 .
\end{aligned}
\end{equation}
\item [b)] (strongly-convex problems) If $f$ is also $\mu$-strongly convex, then
\begin{equation}\label{eqn:distCSGD_SC}
\begin{aligned}
\mathbf{E}\left(  f(\bar x^k) - f(x^\star) \right)  
& \leq \frac{1}{k+1} \frac{1}{2\gamma}\frac{1}{1-3L\gamma}\| x^0 - x^\star\|^2\\
&\hspace{0.4cm} +\frac{3}{1-3L\gamma}\gamma\sigma^2+ \frac{1}{2}\frac{1/\mu+3\gamma}{1-3L\gamma}\epsilon^2,
\end{aligned}
\end{equation}
where $\bar x^k = \sum_{l=0}^k x^l/(k+1)$.
\end{itemize}

\end{theorem}
\begin{proof}
See Appendix \ref{app:CGD}
\end{proof}

Theorem \ref{thm:distCSGD_convex} establishes a sub-linear convergence of D-CSGD toward the optimum with \added{a} residual error depending on \added{the} stochastic gradient noise $\sigma$, compression $\epsilon$, problem parameters $\mu,L$ and the step-size $\gamma.$ In particular, the residual error consists of two terms.
 The first term  comes from the stochastic gradient noise $\sigma^2$ and decreases in  proportion to the step-size. 
 The second term arises from the precision of the compression $\epsilon$, and cannot diminish towards zero no matter how small we choose the step-size. 
 In fact, it can be bounded by noting that
$$   \frac{1+3L\gamma}{1-3L\gamma}>1  ~~~\text{and}~~~ \frac{1}{2}\frac{1/\mu+3\gamma}{1-3L\gamma}>\frac{1}{2\mu}, $$
for all $\gamma\in (0,1/(3L))$. 
 This means that the  upper bound  in Equation~\eqref{eqn:distCSGD_NC} cannot become smaller than $\epsilon^2$
 and the upper bound  in Equation~\eqref{eqn:distCSGD_SC} cannot become smaller than $\epsilon^2/(2\mu)$. 


\subsection{Limits of Direct Compression: Lower Bound} \label{subsec:Limits}
%
%


\replaced{We now show that the bounds derived above are tight.}{
We now show that the bounds provided in Theorem~\ref{thm:CGDQuadratic} and~\ref{thm:distCSGD_convex}  are tight. }
%
 %
 %
\deleted{In particular, we provide a simple example where the $\epsilon$-compressor and standard gradient descent  satisfy the worst-case upper bounds with equality.}

\begin{example}\label{ex:SimpleQ}
\textit{
Consider the scalar optimization problem
\begin{equation*}
\begin{aligned}
& \underset{x}{\text{minimize}}
& &  \frac{\mu}{2} x^2.
\end{aligned}
\end{equation*}
 \replaced{and the iterates generated by the CGD algorithm}{Suppose that we run the compressed gradient decent algorithm:}
 \begin{align} \label{eq:quant_gradient}
   x^{k+1}=x^k-\gamma Q(f'(x^k))=x^k-\gamma\mu Q(x^k),
 \end{align}
 where $Q(\cdot)$ is the $\epsilon$-compression (see Definition~\ref{def:DBEC})
 $$Q(z)= \begin{cases} z-\epsilon \frac{z}{|z|} &\text{if } z\neq 0 \\ \epsilon & \text{otherwise}. \end{cases}  $$ 
If \deleted{we  take} $\gamma\in (0,1/\mu]$  and $|x^0|>\epsilon$ then for all $k\in \mathbb{N}$ we have
 \begin{align*}
    |x^{k+1}-x^{\star}|=|x^{k+1}|=&|x^k-\gamma Q(f'(x^k))|\\
     =&  (1-\mu\gamma) |x^{k}|+\gamma\epsilon \\
     =&  (1-\mu\gamma)^{k+1} |x^0| + \epsilon\gamma \sum_{i=0}^k (1-\mu\gamma)^i \\
     =&  (1-\mu\gamma)^{k+1} |x^0| + \epsilon \gamma \frac{1-(1-\mu\gamma)^{k+1}}{\mu\gamma} \\     
     =&  (1-\mu\gamma)^{k+1}(|x^0|-\epsilon)+ \epsilon/\mu \\
     \geq& \epsilon/\mu,
 \end{align*}
 where we have used that $x^{\star}=0$. In addition, 
 \begin{align*}
 f(\bar x^k) - f(x^\star) &= \frac{\mu}{2}\frac{1}{k+1}\sum_{i=0}^k |x^i|^2 \\ 
 & \geq \frac{1}{2\mu}\epsilon^2,
 \end{align*}  
 where $\bar x^k = \sum_{i=0}^k x^i/(k+1).$
 %
%
%
 }

 \end{example}
 The above example shows that  the $\epsilon$-compressor cannot achieve accuracy better than $\epsilon/\mu$ and $\epsilon^2/(2\mu)$ in terms of  $\| x^k - x^\star \|^2$ and $f(\bar x^k) - f(x^\star)$, respectively. These lower\replaced{ }{-}bounds match the upper\replaced{ }{-}bound in Theorem \ref{thm:CGDQuadratic}, and the upper\replaced{ }{-}bound \deleted{in Equation} \eqref{eqn:distCSGD_SC} in Theorem \ref{thm:distCSGD_convex} if the step-size is sufficiently small. However, in this paper we show the surprising fact that an arbitrarily  good solution accuracy can be obtained 
 with $\epsilon$-compressor and any $\epsilon>0$  if we include a simple correction step in the optimization algorithms.
 

\deleted{
We show that the error compensation scheme indeed reaches lower-than-$\epsilon$ solution accuracy in the next section.
}


\section{Error Compensated Gradient Compression} \label{sec:ECGC}



In this section we illustrate how we can avoid the accumulation of \replaced{compression errors in gradient-based optimization.}{quantization from the direction gradient compression error with a smart error compensation.}   In subsection~\ref{subsec:ECGC-AII}, we \replaced{introduce our error compensation mechanism and illustrate its powers on quadratic problems. }{illustrate the main intuition of error compensated gradient methods on  quadradic problems.}  In subsection~\ref{subsec:ECGC-PG}, we provide \replaced{a more}{the} general error-compensation algorithm and \added{derive} associated convergence results. 
In subsection~\ref{Sec:EC-CA} we discuss the complexity of the algorithm and how it can be reduced with Hessian approximations.

\subsection{Error Compensation: Algorithm and Illustrative Example}\label{subsec:ECGC-AII}
%
%
%
%
%
%
%
{\color{black}
To motivate our algorithm design and demonstrate its theoretical advantages compared to existing methods we first consider quadratic problems with 
%
$$   f(x)=\frac{1}{2}x^T Hx + b^Tx. $$

 The goal of error compensation is to remove compression errors that is accumulated over time.  For the quadratic problem we can design the error compensation ``optimally'' in the sense that it removes all accumulated errors. The iterations of the proposed algorithm (explained below) can be written as  
\begin{equation}\label{eqn:ECCGD}
\begin{aligned}
	x^{k+1} & =x^k - \gamma Q( \nabla f(x^k) + A_\gamma e^k) \\
	e^{k+1} &=  \underbrace{\nabla f(x^k) + A_\gamma e^k}_{\text{Input to Compressor}} - \underbrace{Q( \nabla f(x^k) + A_\gamma e^k)}_{\text{Output from Compressor}}.
\end{aligned}
\end{equation}
with $e^0=0$ and $A_\gamma = I -\gamma H$. 
%
%
%
%
%
%
%
This algorithm is similar to the direct gradient compression in Equation~\eqref{eqn:CGD}.  The main difference is that we have introduced the memory term $e^k$ in the gradient update. 
The term $e^k$ is essentially the \emph{compression error}, the difference  between the compressor input and output. 
%
%
To see why this error compensation removes all accumulated errors we will formulate the algorithm as a linear system.}
To that end, we define the  \emph{gradient error}
$$c^k  = \underbrace{ \nabla f(x^k)}_{\text{True Gradient}} -  \underbrace{Q(\nabla f(x^k) + A_\gamma e^k)}_{\text{Approximated Gradient Step}} .$$
and re-write the evolution of the compression error as
\begin{align*}
e^{k+1} = c^k + A_\gamma e^k. 
\end{align*}
This relationship implies that 
\begin{align*}
  \quad e^k = \sum_{j=0}^{k-1} A^{k-1-j}_\gamma c^j.
\end{align*}
With this in mind, we can \added{re-}write the $x$-update \replaced{as}{equivalently, similarly as in Equation~\eqref{eq:xupdate}, as} 
 \begin{align*}
      x^{k+1}=A_{\gamma}x^k-\gamma b+ \gamma c^k
 \end{align*}
 \replaced{and establish that}{which reduces to} 
%
\begin{align}
x^{k+1} - x^\star  & = A^{k+1}_\gamma (x^0 - x^\star) +  \gamma \sum_{i=0}^{k} A^{k-i}_\gamma c^i  \notag \\
							& = A^{k+1}_\gamma (x^0 - x^\star) +  \gamma e^{k+1}. \label{eq:LinSys}
\end{align} 
Note that in contrast to Equation~\eqref{eq:FGC-QD}, the residual error now only depends on the latest compression error $e^{k+1}$, and no longer of the accumulated past compression errors. 
  In particular,   if $Q(\cdot)$ is an $\epsilon$-compressor then $||e^{k+1}||\leq \epsilon$ and we have a constant upper bound on the error. This means that we can recover high solution accuracy by proper tuning of the step-size.  We illustrate this in the following theorem. 


\begin{theorem}\label{thm:ErrorCompensatedSC}
Consider the quadratic optimization problem \replaced{with}{over the} objective function~\eqref{eq:quadOBJ} where $H$ is positive definite, and let $\mu$ and $L$ be the smallest and largest eigenvalues of $H$, respectively.  Then, the iterates $\{x^k\}_{k\in\mathbb{N}}$ generated by \eqref{eqn:ECCGD} with $A^k=I-\gamma H$ and $e^0=0$ satisfy
\begin{align*}
    \| x^k - x^\star  \|  \leq \rho^k\| x^0 - x^\star \| + \gamma \epsilon, 
\end{align*}
where 
\begin{align*}
\rho &= \left\{ \begin{array}{ll} 1-1/\kappa & \text{if } \quad \gamma = 1/L \\   1-{2}/{(\kappa+1)} & \text{if }\quad \gamma = 2/(\mu+L)   \end{array}\right.  ,
\end{align*}
and $\kappa = L/\mu$.
\end{theorem}
\begin{proof}
See Appendix \ref{app:thm:ErrorCompensatedSC}.
\end{proof}
Theorem \ref{thm:ErrorCompensatedSC} implies that error-compensated gradient descent has linear convergence rate and \replaced{can attain}{recover} arbitrarily high solution accuracy \replaced{by decreasing the step-size.}{. The quantization error term can \replaced{be made arbitrarily small by decreasing the step-size.}{decrease toward zero by diminishing step-size.}}   \replaced{Comparing with Theorem~\ref{thm:CGDQuadratic}, we note that  error compensation attains lower residual error than direct compression if we insist on maintaining the same convergence rate.}{In addition,  error compensation achieves the same rate as but lower residual error than direct gradient compression as shown in Theorem \ref{thm:CGDQuadratic}.} In particular, error compensation in Equation \eqref{eqn:ECCGD} with $\gamma=1/L$ and $\gamma=2/(\mu+L)$ reduces compression error $\kappa$ and $(\kappa+1)/2$, respectively. \added{ Hence, the benefit is especially pronounced for} ill-conditioned \deleted{quadratic optimization} problems~\cite{ourICASSPworks}. 
%
\textcolor{black}{We illustrate this theoretical advantage of our error-compensation compared existing schemes in the following remark.
}


{\color{black}
\begin{remark}[Comparison to existing schemes] \label{remark:Comparison}
 Existing error-compensations for compressed gradients keep in memory the sum (or weighted sum) of all previous compression errors~\cite{alistarh2018convergence,wu2018error,Stich2018,2019arXiv190109847P,tang2019doublesqueeze,tang2019texttt}. 
 We can express this here  by changing the algorithm in Eq.~\eqref{eqn:ECCGD} to
\begin{equation}\label{eqn:ECCGD_Vr1}
\begin{aligned}
	x^{k+1} & =x^k - \gamma Q( \nabla f(x^k) + \alpha e^k) \\
	e^{k+1} &=  \underbrace{\nabla f(x^k) + \alpha e^k}_{\text{Input to Compressor}} - \underbrace{Q( \nabla f(x^k) + \alpha e^k)}_{\text{Output from Compressor}}.
\end{aligned}
\end{equation}
\replaced{with}{where} $e^0=0$ and $\alpha\in (0,1]$. 
If we perform a similar convergence study as above (cf. Eq.~\eqref{eq:LinSys}) then we get
\begin{align*}
    x^{k+1} - x^\star & = A_\gamma^{k+1}(x^0-x^\star) + \gamma e^{k+1} + \gamma \sum_{l=0}^{k} A_{\gamma}^{k-l} B_{\alpha,\gamma} e^l 
\end{align*}
 where $A_\gamma = I - \gamma H$ and $B_{\alpha,\gamma} = (1-\alpha) I - \gamma H$.  The final term shows that these error compensation schemes do not remove the accumulated quantization errors, even though they have been shown to outperform direct compression. However, our error compensation does remove all of the accumulated error, as shown in Eq.~\eqref{eq:LinSys}. This shows why second-order information improves the accuracy of error compensation.
 \end{remark}}

\subsection{Partial Gradient Communication}   \label{subsec:ECGC-PG}


For \deleted{general} optimization with partial gradient communication\added{,} the natural generalization of error-compensated gradient algorithms \replaced{consist of the following steps:}{is to follow the steps} at 
each iteration in parallel\added{,}  worker nodes compute their local stochastic gradients $g_i(x;\xi_i)$ and \replaced{add a local error compensation term $e_i$ before applying the $\epsilon$-compressor}{compress them which accounts for the memory $e_i$}. \replaced{The}{Then, the} master node waits for all compressed gradients and \replaced{updates}{maintains} the decision vector by \deleted{performing the gradient descent step:}
\begin{equation}\label{eqn:DECCSGD}
\begin{split}
x^{k+1} &= x^k - \gamma\frac{1}{n}\sum_{i=1}^n Q(g_i(x^k;\xi_i^k) + A_i^k e_i^k), 
\end{split}
\end{equation} 
while each worker $i$ \replaced{updates its}{maintains the} memory $e_i$ according to
\begin{align} \label{eqn:DECCSGD-e}
e_{i}^{k+1} & = g_i(x^k;\xi_i^k) + A_i^k e_i^k - Q(g_i(x^k;\xi_i^k) + A_i^k e_i^k).
\end{align}
Similarly as in the previous subsection, we define\footnote{\textcolor{black}{As discussed in Remark~\ref{remark:Comparison}, existing error-compensation schemes are recovered by setting $A_i^k =  \alpha I$ where $\alpha\in(0,1]$.}}
\begin{equation} \label{eqn:defOfAik}
      A_i^k= I - \gamma H_i^k
\end{equation}
where $H_i^k$ is either  a deterministic or  stochastic version of the Hessian $\nabla^2 f_i(x^k)$. 
In this paper, we define the stochastic Hessian  in analogus way as the stochastic gradient as follows:
\begin{align}
&\mathbf{E} [H_i^k ] = \nabla^2 f_i(x^k), \quad \text{and}  \label{eqn:unbiasedHessian}\\ 
&\mathbf{E} \|  H_i^k - \nabla^2 f_i(x^k)\|^2 \leq \sigma_H^2. \label{eqn:boundedVRHessian}
\end{align}
Notice that $H_i^k$ is a local information of worker $i$. In real implementations, each worker can form  the stochastic Hessian and the stochastic gradient independently at random. 
 %
%
%
%
%
%
{\color{black}
In the deterministic case the algorithm has similar convergence properties as the error compensation for the quadradic problems \replaced{studied above.}{in the previous section (Theorem  \ref{thm:ErrorCompensatedSC}).}
\begin{theorem} \label{thm:5}
 Consider the optimization problem \eqref{eqn:Problem} where $f_i(\cdot)$ are $L$-smooth and $f(\cdot)$ is $\mu$-strongly convex. Suppose that $g_i(x^k;\xi_i^k)=\nabla f_i(x^k)$ and $H_i^k = \nabla^2 f_i(x^k)$. If $Q(\cdot)$ is the $\epsilon$-compressor and $\gamma = 2/(\mu+L)$ then 
     \begin{align*}
        \|x^k - x^\star \| \leq \rho^k \|x^0 - x^\star \| + \gamma \epsilon C,
    \end{align*}
    where $\rho = 1 - 2/(\kappa+1)$, $\kappa=L/\mu$, $C=1+\gamma L(\kappa+1)$.
\end{theorem}
\begin{proof}
  See Appendix~\ref{App:Pr5}.
\end{proof}
 The theorem shows that the conclusions from the quadratic case can be extended to general strongly-convex functions and multiple nodes. In particular, the algorithm converges linearly to an approximately optimal solution with higher precision as the step-size decreases. 
 We now illustrate the results in the stochastic case.
}

%
%
%
%
%
%
\begin{theorem}\label{thm:distEC-CSGD_convex}
Consider the optimization problem \eqref{eqn:Problem} where each $f_i(\cdot)$ is $L$-smooth, and the iterates $\{x^k\}_{k\in\mathbb{N}}$ generated by \eqref{eqn:DECCSGD} with $A_i^k$ defined by Equation \eqref{eqn:defOfAik}, under the assumptions of stochastic gradients $g_i(x^k; \xi_i^k)$ in Equation~\eqref{eqn:unbiasednessSGD} and \eqref{eqn:boundedvarianceSGD}, and stochastic Hessians $H_i^k$ in Equation~\eqref{eqn:unbiasedHessian} and \eqref{eqn:boundedVRHessian}.
%
Assume that $Q(\cdot)$ is an $\epsilon$-compressor and that $e^0_i=0$ for all $i\in[1,n]$. 
\begin{itemize}
\item [a)]  (non-convex problems) If $\gamma<1/(3L)$, then 
\begin{align*}
\mathop{\min}\limits_{l\in[0,k]} \mathbf{E} \| \nabla f(x^l) \|^2 
& \leq \frac{1}{k+1}\frac{2}{\gamma}\frac{1}{1-3L\gamma}(f(x^0)-f(x^\star)) \\ 
&\hspace{0.4cm}  + \frac{3L}{1-3L\gamma}\gamma\sigma^2+ \frac{\alpha_2}{1-3L\gamma}\gamma^2\epsilon^2 ,
\end{align*}
where $\alpha_2 = L^2 + (2+6L\gamma)(\sigma_H^2+L^2)$.

\item [b)] (strongly-convex problems) If  $f$ is also $\mu-$strongly convex, and \deleted{moreover} $\gamma<  (1-\beta)/(3L) $ \replaced{with}{where} $0<\beta<1$, then 
\begin{align*}
\mathbf{E}\left(f(\bar x^k) - f(x^\star) \right)
 & \leq \frac{1}{k+1}\frac{1}{2\gamma}\frac{1}{1-\beta-3L\gamma}\| x^0  - x^\star \|^2 \\ 
 &\hspace{0.4cm} + \frac{3}{2}\frac{1}{1-\beta-3L\gamma}\gamma \sigma^2 \\
  & \hspace{0.4cm} + \frac{1}{2}\frac{\alpha_1}{1-\beta-3L\gamma}\gamma^2\epsilon^2, 
\end{align*}
where $\alpha_1=\mu+L/\beta + \left({4}/{\mu}+6\gamma\right)(\sigma_H^2+ L^2)$ and $\bar x^k = \sum_{l=0}^k x^l/(k+1)$.

\end{itemize}

\end{theorem}
\begin{proof}
See Appendix \ref{app:ECCGD}.
\end{proof}
%
%
%
%
%
%
The theorem establishes that our  error-compensation method converges with rate $\mathcal{O}(1/k)$ toward the optimum with \added{a} residual error. 
 Like Theorem~\ref{thm:distCSGD_convex} for direct gradient compression, 
 the residual error consists of two terms.
 The first residual term depends on the stochastic gradient noise $\sigma^2$ and the second term depends on the precision of the compression $\epsilon$. 
 The first term can be made arbitrary small by decreasing the step-size $\gamma>0$, similarly as in Theorem~\ref{thm:distCSGD_convex}. 
 However, unlike in Theorem~\ref{thm:distCSGD_convex}, here we can make the second residual term arbitrarily small by decreasing $\gamma>0$. 
 In particular, for a fixed $\epsilon>0$,  the second residual term goes to zero at the rate $\mathcal{O}(\gamma^2)$.  
 \textcolor{black}{We can thus get an arbitrarily high solution accuracy even when the compression resolution $\epsilon$ is fixed. However, the cost of increasing the solution accuracy by decreasing $\gamma$ is that it slows down the  convergence (which is proportional to $1/\gamma$).} 

We validate the superior performance of Hessian-based error compensation over \deleted{the} existing \replaced{schemes}{counterpart} in Section~\ref{sec:numericalresults}. To \replaced{reduce computing and memory requirements}{save Hessian computations}, we \deleted{also} propose \replaced{a Hessian}{the diagonal} approximation (\deleted{i.e. }\replaced{using}{keeping} only the diagonal elements of the Hessian)\deleted{ which leads to \replaced{a}{the} computation cost of $\mathcal{O}(d)$}. \replaced{Error compensation with this approximation}{This diagonal Hessian error compensation} is shown to have comparable performance to \replaced{using the full Hessian.}{its full Hessian-based counterpart. }

{\color{black}
\subsection{Algorithm Complexity \& Hessian Approximation}\label{Sec:EC-CA}

Our scheme improves the iteration complexity of compressed gradient methods, both of the methods that use direct compression and error-compensation.  
 This reduces the number of  gradient transmissions, which in turn makes our compression more communication efficient than existing schemes.  However, the improved communication complexity comes at the price of additional computations, since our compression uses second-order information.  In particular, from Equations~(24) and~(25),  computing the compressed gradient at each node requires $\mathcal{O}(d^2)$ arithmetic operations to multiply the  Hessian matrix by the compression error.  
    On the other hand,  direct compression and existing compensation methods require only $\mathcal{O}(d)$ operations to compute the compressed gradient. 
  Thus, our compression is more communication efficient than existing schemes but achieves that by additional computations at nodes each iteration. 
  We can improve the computational efficiency of our error-compensation by using computationally efficient Hessian approximations. 
   For example, the Hessian can be approximated by using only its diagonal elements.  This reduces the  computation of each compression to $\mathcal{O}(d)$ operations, comparable to existing schemes. We show in the next section that this approach gives good results on both convex and non-convex problems. 
   Alternatively, we might use standard Hessian approximations from quasi-Newton methods such as BFGS~[31],~[32] or update   the Hessian less frequently.

}



\section{Numerical Results}\label{sec:numericalresults}

In this section, we validate the superior convergence properties of Hessian-aided error compensation compared to the state-of-the-art. We also show that error compensation with a diagonal Hessian approximation shares many benefits with the full Hessian algorithm. 
In particular, we evaluate the error compensation schemes on centralized SGD and distributed gradient descent for \eqref{eqn:Problem} with component functions on the form \eqref{eqn:fi_ERM} and $\lambda=0$.
%
%
\textcolor{black}{
In all simulations, we normalized each data sample by its Euclidean norm and used the initial iterate $x^0={0}$. In plot legends,   {\fontfamily{qcr}\selectfont Non-EC} denotes the compressed gradient method \eqref{eqn:QDGD}, while {\fontfamily{qcr}\selectfont EC-I},  {\fontfamily{qcr}\selectfont EC-H} and  {\fontfamily{qcr}\selectfont EC-diag-H}  are error compensated methods 
governed by the iteration described in Equation \eqref{eqn:DECCSGD} with $A_i^k = I$, $A^k_i=I-\gamma H^k_i$ and  $A^k_i=I-\gamma \text{diag}(H^k_i)$, respectively. Here,  $H^k_i$ is the Hessian information matrix associated with the stochastic gradient $g_i(x^k;\xi^k_i)$ and $\text{diag}(H^k_i)$ is a matrix with the diagonal entries of $H^k_i$ on its diagonal and zeros elsewhere.  Thus, {\fontfamily{qcr}\selectfont EC-I} is the existing state-of-the-art error compensation scheme in the literature, {\fontfamily{qcr}\selectfont EC-H} denotes our proposed Hessian-aided error compensation, and {\fontfamily{qcr}\selectfont EC-diag-H} is the same error compensation using a diagonal Hessian approximation.
}

\begin{figure}[htb]
    \centering 
\includegraphics[width=0.8\linewidth]{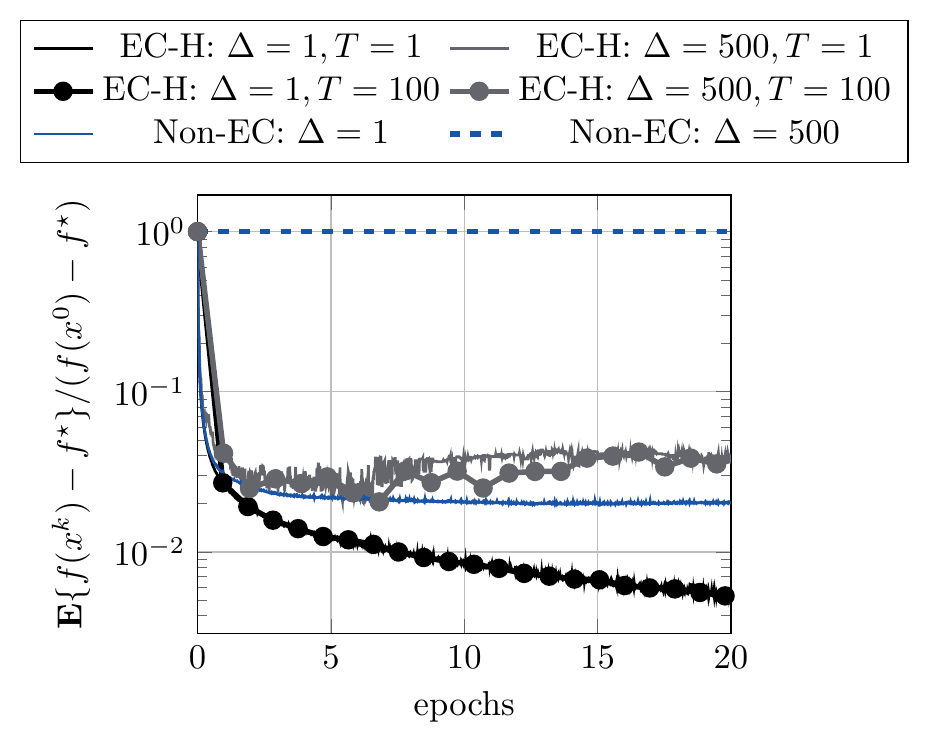}
\caption{Comparisons of D-CSGD and D-EC-CSGD with one worker node using different compensation schemes for the least-squares problems over {\fontfamily{qcr}\selectfont a3a}  from \cite{chang2011libsvm} when the deterministic rounding quantizer is applied. We set the step-size $\gamma=0.1/L$, the mini-batch size  $b=\vert\mathcal{D}\vert/20$, where  $\vert\mathcal{D}\vert$ is the total number of data samples.}
\label{fig:compressedSGD_a3a_vary_m_precision}
\end{figure}

\begin{figure*}[htb]
    \centering 
\begin{subfigure}{0.3\textwidth}
  \includegraphics[width=\linewidth]{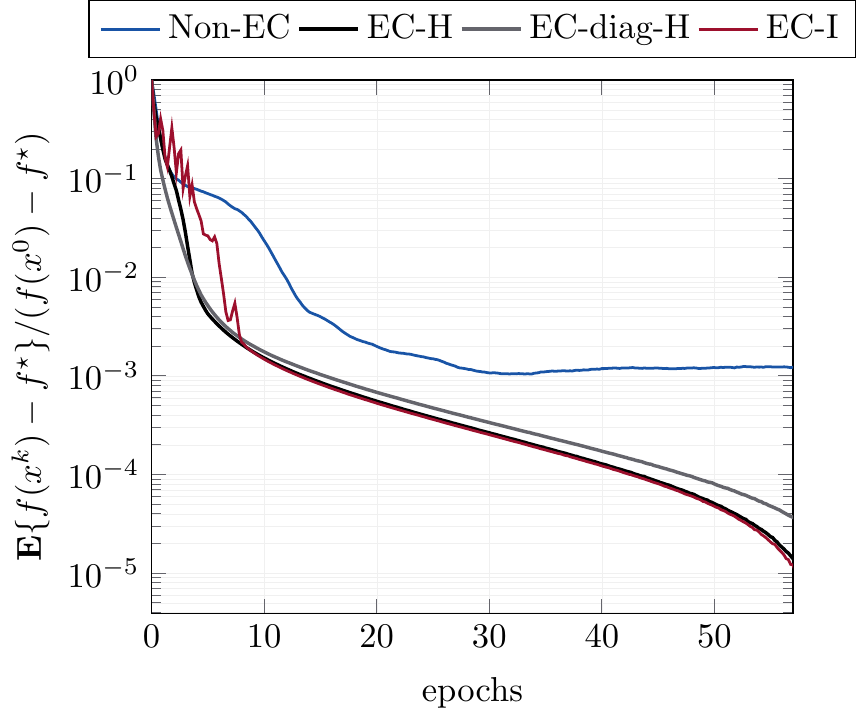}
  \caption{{\fontfamily{qcr}\selectfont mushrooms} }
  \label{fig:1}
\end{subfigure}\hfil 
\begin{subfigure}{0.3\textwidth}
  \includegraphics[width=\linewidth]{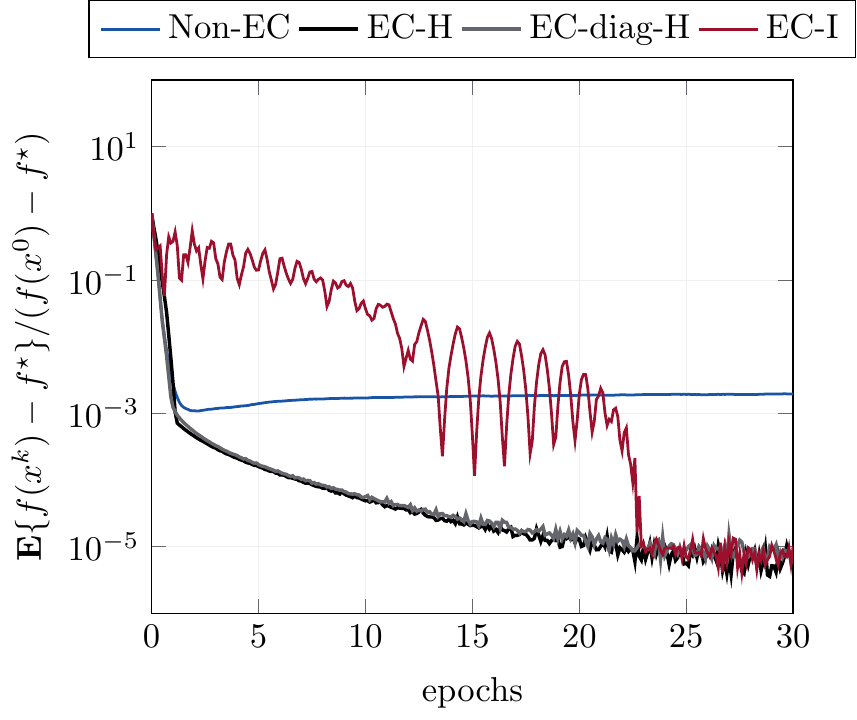}
  \caption{{\fontfamily{qcr}\selectfont a9a} }
  \label{fig:2}
\end{subfigure}\hfil 
\begin{subfigure}{0.3\textwidth}
  \includegraphics[width=\linewidth]{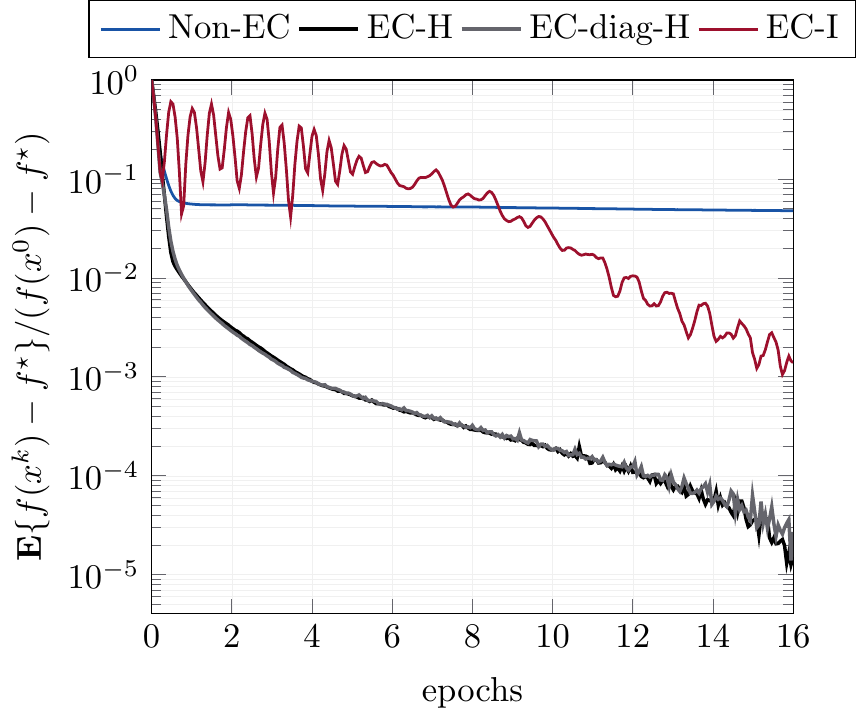}
  \caption{{\fontfamily{qcr}\selectfont w8a} }
  \label{fig:CSGD}
\end{subfigure}

\caption{Comparisons of D-CSGD and D-EC-CSGD with one worker node  using different compensation schemes for the least-squares problems over bench-marking data sets from \cite{chang2011libsvm} when the binary (sign) compression is applied. We set the step-size $\gamma=0.1/L$, and the mini-batch size $b=\vert\mathcal{D}\vert/10$, where  $\vert\mathcal{D}\vert$ is the total number of data samples.}
\label{fig:compressedSGD}
\end{figure*}

\begin{figure*}[htb]
    \centering 
\begin{subfigure}{0.3\textwidth}
  \includegraphics[width=\linewidth]{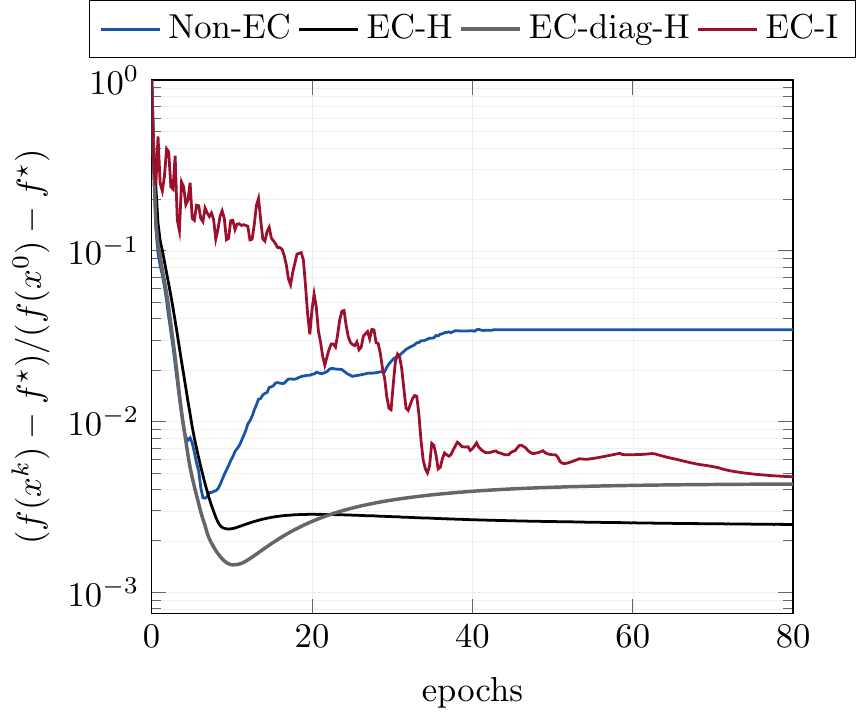}
\caption{{\fontfamily{qcr}\selectfont mushrooms} }
  \label{fig:4}
\end{subfigure}\hfil 
\begin{subfigure}{0.3\textwidth}
  \includegraphics[width=\linewidth]{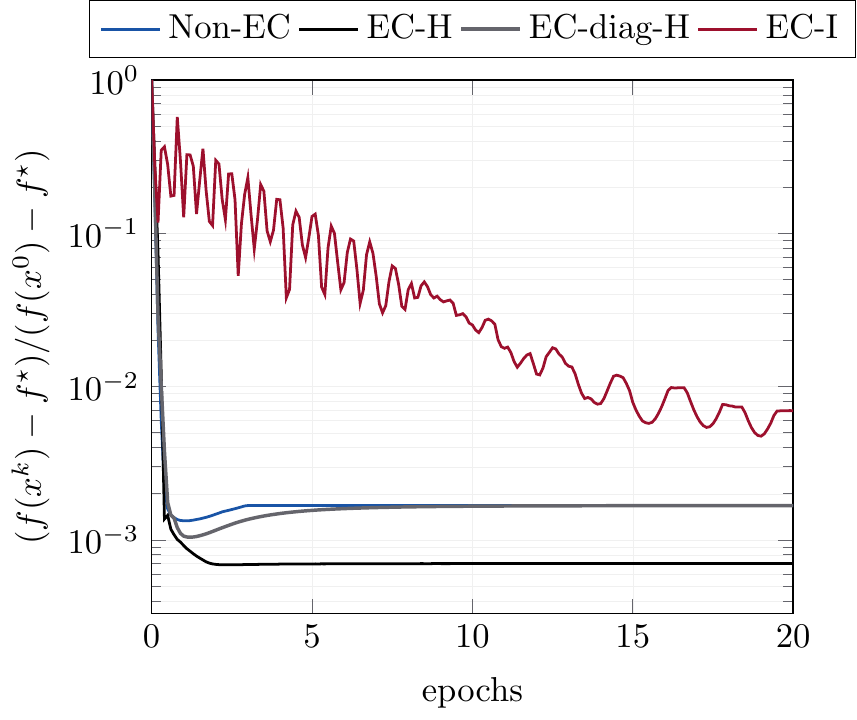}
\caption{{\fontfamily{qcr}\selectfont a9a} }
  \label{fig:5}
\end{subfigure}\hfil 
\begin{subfigure}{0.3\textwidth}
  \includegraphics[width=\linewidth]{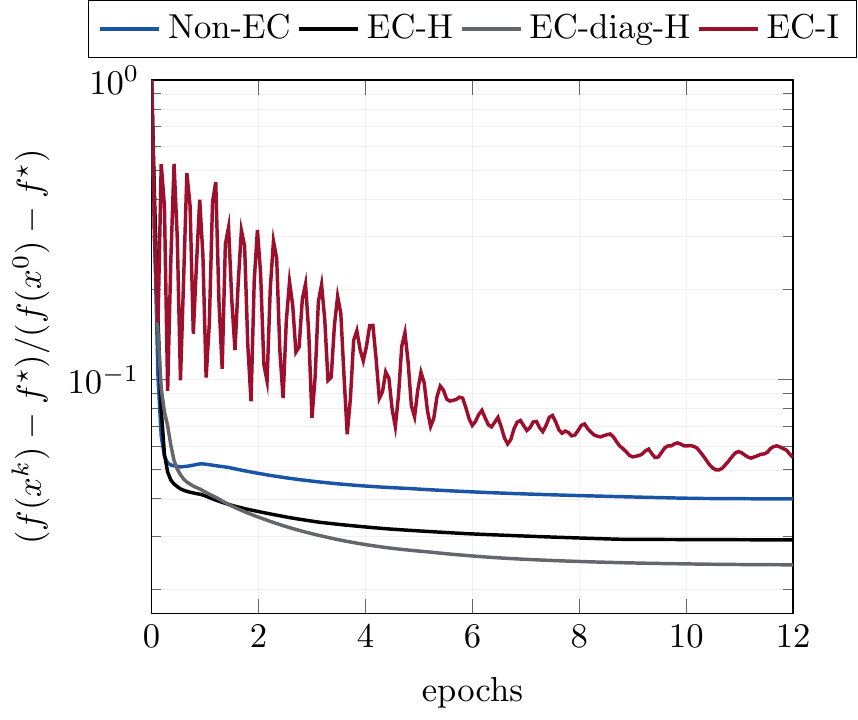}
 \caption{{\fontfamily{qcr}\selectfont w8a} }
  \label{fig:6}
\end{subfigure}

\caption{Comparisons of D-CSGD and D-EC-CSGD with $\nabla g_i(x;\xi_i)=\nabla f_i(x)$ using different compensation schemes for the least-squares  problems over bench-marking data sets from \cite{chang2011libsvm} when the binary (sign) compression is applied. We set  the step-size $\gamma=0.1/L$, and $5$ worker nodes. }
\label{fig:distGD}
\end{figure*}

{
\color{black}
\subsection{Linear Least Squares Regression}

We consider the least-squares regression problem \eqref{eqn:Problem} with each component function on the form \eqref{eqn:fi_ERM}, with $\lambda=0$ and  $$\ell(x; z_i^j,y_i^j)=  (\langle z_i^j , x \rangle - y_i^j)^2/2.
$$ 
Here, $(z_i^1,y_i^1),\ldots,(z_i^m,y_i^m)$ are its data samples with feature vectors $z_i^1\in\mathbb{R}^d$ and associated class labels $y_i^j\in\{-1,1\}$.
Clearly, $f_i(\cdot)$ is strongly convex and smooth with parameters $\mu_i$ and $L_i$, denoting the smallest and largest eigenvalues, respectively, of the matrix
$
A_i = \sum_{j=1}^m z_i^j (z_i^j)^T.
$
Hence, this problem has an objective function $f(\cdot)$ which is $\mu$-strongly convex and $L$-smooth with $\mu=\min_{i\in[1,n]} \mu_i$ and $L=\max_{i\in[1,n]} L_i$, respectively.

We evaluated full gradient methods with three Hessian-based compensation variants; {\fontfamily{qcr}\selectfont EC-H}, {\fontfamily{qcr}\selectfont EC-diag-H} and {\fontfamily{qcr}\selectfont EC-BFGS-H}. Here, {\fontfamily{qcr}\selectfont EC-BFGS-H} is the compensation update where the full Hessian  $H^k$ is approximated using the BFGS update \cite{nocedal2006numerical}. Figure \ref{fig:compressedGD_bounds_BFGS} suggests that the worst-case bound from Theorem \ref{thm:ErrorCompensatedSC}  is tight for error compensated methods with {\fontfamily{qcr}\selectfont EC-H}.
In addition, compensation updates that approximate Hessians by using only the diagonal elements and by the BFGS method perform worse than the full Hessian compensation scheme.

\begin{figure}[htb]
    \centering 
\includegraphics[width=0.8\linewidth]{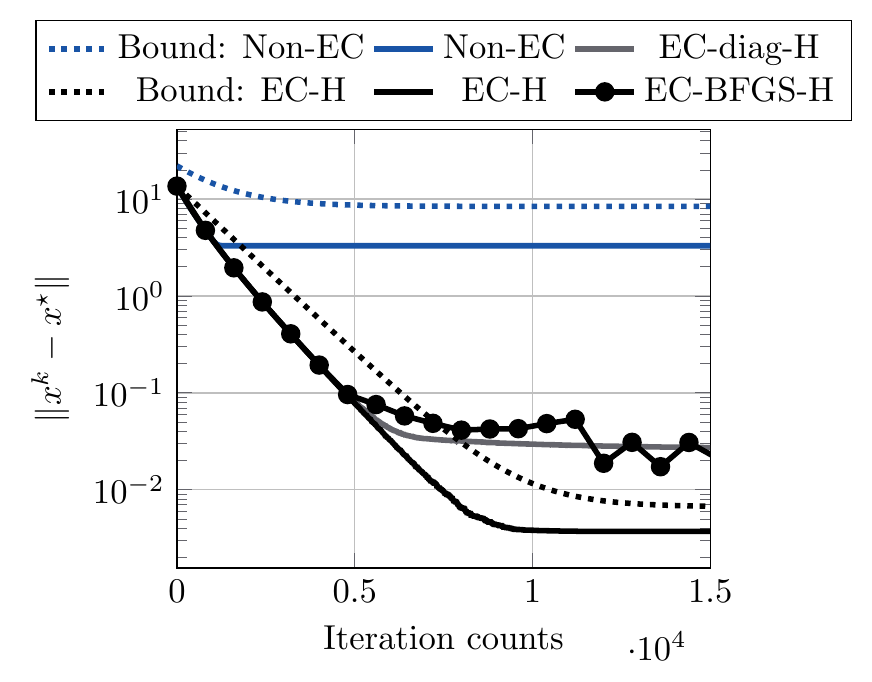}
\caption{Comparisons of D-CSGD and D-EC-CSGD with one worker node and with $g_i(x;\xi_i)=\nabla f_i(x)$ for the least-squares problems over synthetic data with $4,000$ data points and $400$ features, when the deterministic rounding quantizer with $\Delta=1$ is applied. We set the step-size $\gamma=2/(\mu+L)$.}
\label{fig:compressedGD_bounds_BFGS}
\end{figure}

We begin by considering the deterministic rounding quantizer \eqref{eqn:RoundingQuant}. 
Figure \ref{fig:compressedSGD_a3a_vary_m_precision} shows that compressed SGD cannot reach a high solution accuracy, and its performance deteriorates as the quantization resolution decreases (the compression is coarser). Error compensation, on the other hand, achieves a higher solution accuracy and is surprisingly insensitive to the amount of compression.

Figures \ref{fig:compressedSGD} and \ref{fig:distGD} evaluate error compensation the binary (sign) compressor on several data sets in both single and multi-node settings.   
Almost all variants of error compensation achieve higher solution accuracy  after a sufficiently large number of iterations. In particular, {\fontfamily{qcr}\selectfont EC-H} outperforms the other error compensation schemes in terms of high convergence speed and low residual error guarantees for centralized SGD and distributed GD. In addition, {\fontfamily{qcr}\selectfont EC-diag-H} has almost the same performance as {\fontfamily{qcr}\selectfont EC-H}. 
}

{\color{black}
\subsection{Non-convex Robust Linear Regression}

Next, we consider the robust linear regression problem \cite{xu2019newton,beaton1974fitting}, with the component function \eqref{eqn:fi_ERM}, $\lambda=0$ and  $$\ell(x; z_i^j,y_i^j)=  (\langle z_i^j , x \rangle - y_i^j)^2/(1+ (\langle z_i^j , x \rangle - y_i^j)^2 ).$$ Here, $f_i(\cdot)$ is smooth with $L_i=\sum_{j=1}^m \| z^j_i \|^2/(6\sqrt{3})$, and thus  $f(\cdot)$ is smooth with the parameter $L=\max_{i\in[1,n]} L_i$. 
}

\begin{figure*}[htb]
    \centering 
\begin{subfigure}{0.31\textwidth}
  \includegraphics[width=\linewidth]{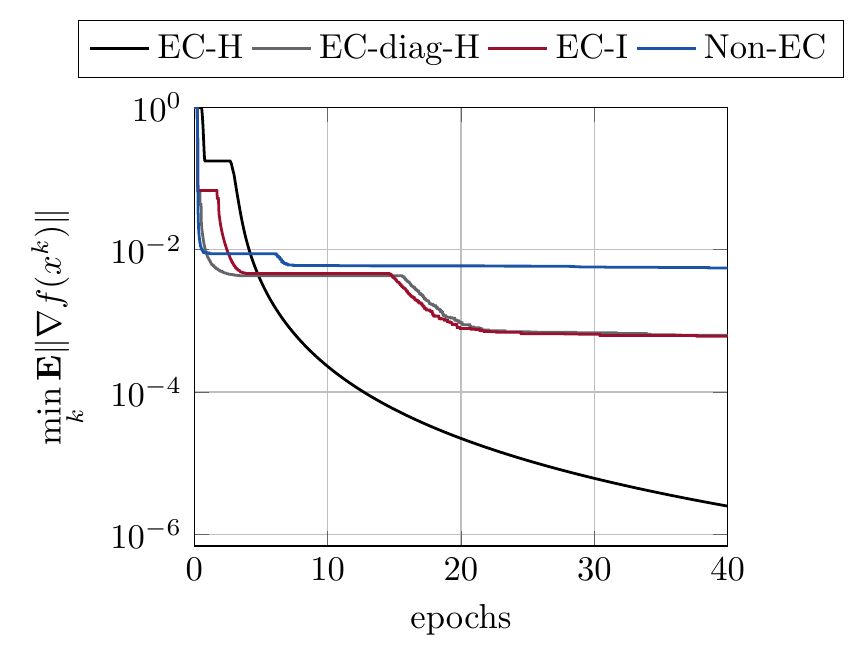}
  \caption{{\fontfamily{qcr}\selectfont a3a} }
  \label{fig:1_NC}
\end{subfigure}\hfil 
\begin{subfigure}{0.31\textwidth}
  \includegraphics[width=\linewidth]{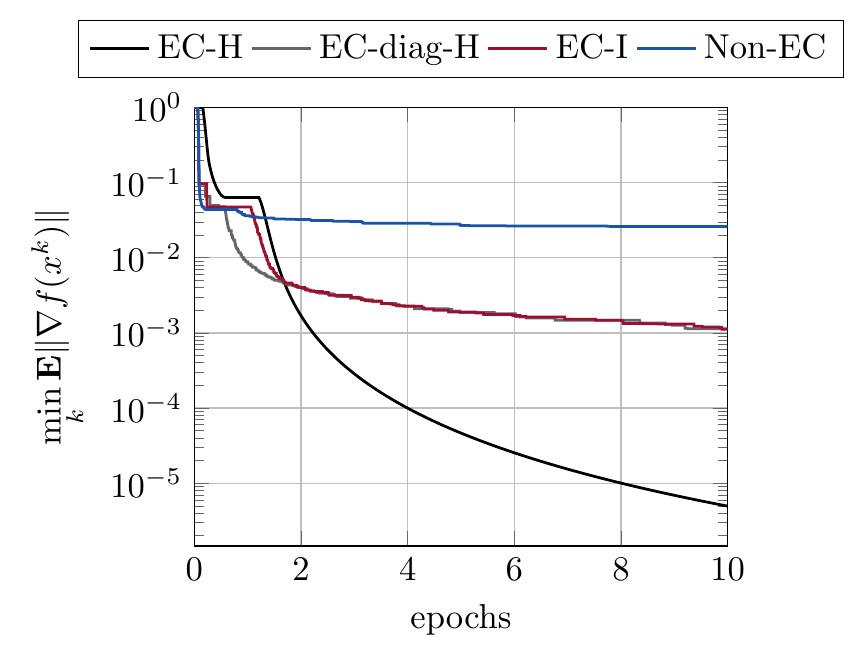}
  \caption{{\fontfamily{qcr}\selectfont mnist} }
  \label{fig:2_NC}
\end{subfigure}\hfil 
\begin{subfigure}{0.31\textwidth}
  \includegraphics[width=\linewidth]{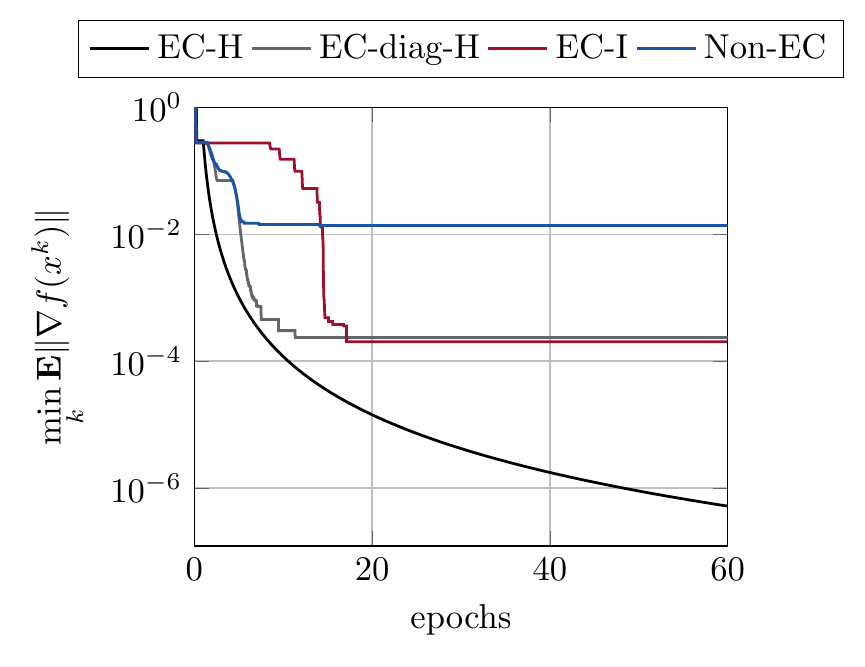}
  \caption{{\fontfamily{qcr}\selectfont phishing} }
  \label{fig:CSGD_NC}
\end{subfigure}

\caption{Comparisons of D-CSGD and D-EC-CSGD with one worker node  using different compensation schemes for non-convex robust linear regression over bench-marking data sets from \cite{chang2011libsvm} when the binary (sign) compression is applied. We set the step-size $\gamma= 1/(60\sqrt{3}L)$, and the mini-batch size $b=\vert\mathcal{D}\vert/10$, where  $\vert\mathcal{D}\vert$ is the total number of data samples.}
\label{fig:compressedSGDNonConvex}
\end{figure*}

\begin{figure*}[htb]
    \centering 
\begin{subfigure}{0.31\textwidth}
  \includegraphics[width=\linewidth]{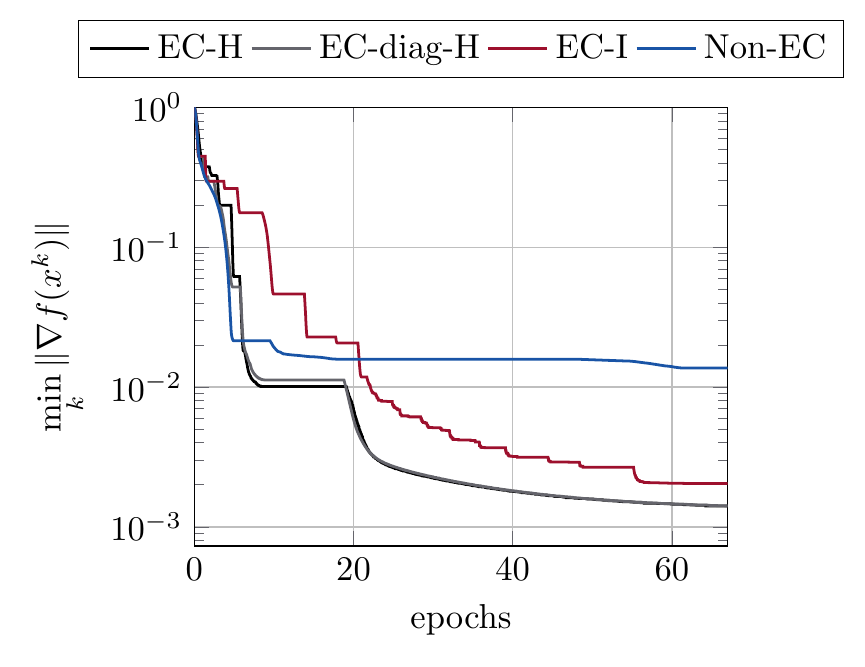}
  \caption{{\fontfamily{qcr}\selectfont a3a} }
  \label{fig:1_dist_NC}
\end{subfigure}\hfil 
\begin{subfigure}{0.31\textwidth}
  \includegraphics[width=\linewidth]{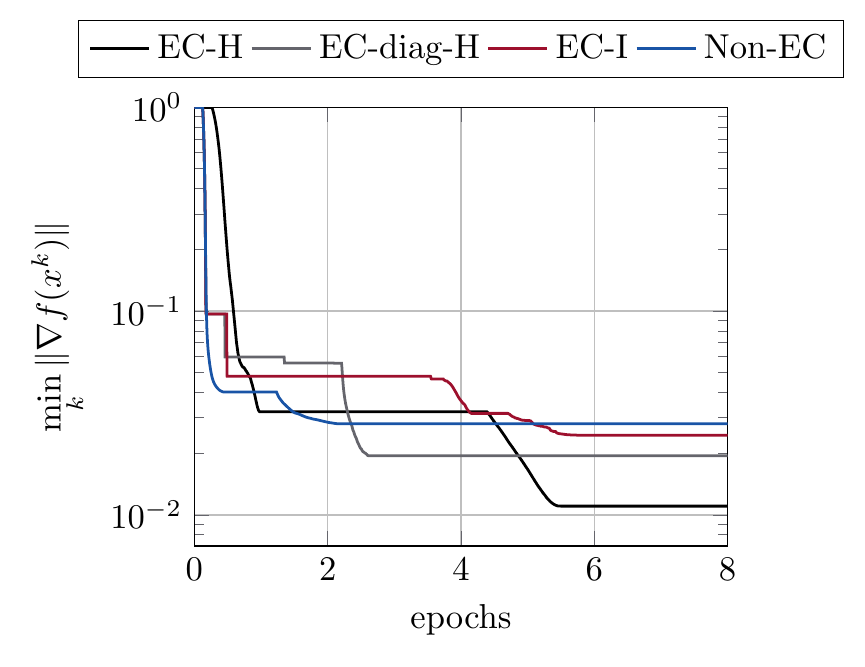}
  \caption{{\fontfamily{qcr}\selectfont mnist} }
  \label{fig:2_dist_NC}
\end{subfigure}\hfil 
\begin{subfigure}{0.31\textwidth}
  \includegraphics[width=\linewidth]{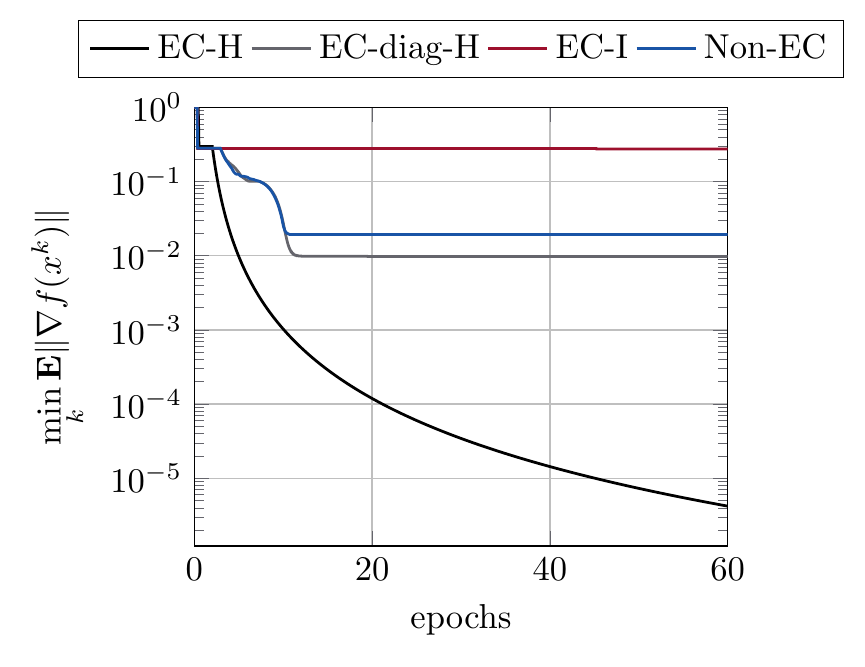}
  \caption{{\fontfamily{qcr}\selectfont phishing} }
  \label{fig:CSGD_dist_NC}
\end{subfigure}

\caption{Comparisons of D-CSGD and D-EC-CSGD with $g_i(x;\xi_i)=\nabla f_i(x)$ using different compensation schemes  for non-convex robust linear regression over bench-marking data sets from \cite{chang2011libsvm} when the binary (sign) compression is applied. We set the step-size $\gamma= 1/(60\sqrt{3}L)$, and 5 worker nodes.}
\label{fig:compresseddistCGDNonConvex}
\end{figure*}

{\color{black}
We consider the binary (sign) compressor and evaluated many compensation algorithms on different data sets; see Figures~\ref{fig:compressedSGDNonConvex} and~\ref{fig:compresseddistCGDNonConvex},
  Compared with direct compression, almost all error compensation schemes improve the solution accuracy, and {\fontfamily{qcr}\selectfont EC-H} provides a higher accuracy solution than {\fontfamily{qcr}\selectfont EC-diag-H} and {\fontfamily{qcr}\selectfont EC-I} for both centralized and distributed architectures.
}



\section{Conclusion}\label{sec:conclusion}

In this paper, we provided a theoretical support for error-compensation in compressed gradient methods.  In particular, we showed that optimization methods with Hessian-aided error compensation can, unlike existing schemes,  avoid \emph{all} accumulated compression errors on quadratic problems and provide accuracy gains on ill-conditioned problems.  We also provided strong convergence guarantees of  Hessian-aided error-compensation for centralized and decentralized stocastic gradient methods on both convex and nonconvex problems.  The superior performance of Hessian-based compensation compared to other error-compensation methods was illustrated numerically on classification problems using large benchmark data-sets in machine learning.  Our experiments showed that error-compensation with diagonal Hessian approximation can achieve comparable performance as the full Hessian while saving the computational costs.


{\color{black}

Future research directions on error compensation include the extensions for federated optimization and the use of efficient Hessian approximations. 
These federated optimization methods communicate compressed decision variables, rather than gradients in compressed gradient methods. 
Error compensation were empirically reported to improve solution accuracy of federated learning methods by initial studies in, \emph{e.g.}, \cite{tang2019texttt,basu2019qsparse}.
However, there is no theoretical justification which highlights the impact of error compensation on compressed decision variables and gradients. 
Another interesting direction is to combine Hessian approximation techniques into the error compensation. For instance, limited BFGS, which requires low storage requirements, can be used in error compensation to approximate the Hessian for solving high-dimensional and even non-smooth problems  \cite{lewis2013nonsmooth,yang2019stochastic,stella2017forward}.  
}


\bibliographystyle{unsrt}
\bibliography{sample.bib}

 \appendices
\section{Review of Useful Lemmas}
This section states  lemmas which are instrumental in our convergence analysis. 
\begin{lemma}\label{lemma:norm_sq_trick}
For $x_i\in\mathbb{R}^d$ and a natural number $N$,
$$\left\| \sum_{i=1}^N x_i \right\|^2 \leq N \sum_{i=1}^N \left\| x_i\right\|^2.$$ 
\end{lemma}

\begin{lemma}\label{lemma:norm_sq_theta_trick}
For $x,y\in\mathbb{R}^d$ and a positive scalar $\theta$,
$$\left\| x+y \right\|^2 \leq  (1+\theta)\| x \|^2 + (1+1/\theta)\|y\|^2.$$
\end{lemma}

\begin{lemma}\label{lemma:innerproduct_theta_trick}
For $x,y\in\mathbb{R}^d$ and a positive scalar $\theta$,
$$2\langle x,y \rangle \leq  \theta\| x \|^2 + (1/\theta)\|y\|^2.$$ 
\end{lemma}
\begin{lemma}[\cite{nesterov2013introductory}]\label{lemma:Nesterov_trick}
Assume that $f$ is convex and $L-$smooth, and the optimimum is denoted by $x^\star$. Then,  
\begin{align}\label{eqn:Nesterov_trick}
\| \nabla f(x)\|^2 \leq 2L (f(x) - f(x^\star)), \quad \text{for } x\in\mathbb{R}^d.
\end{align}
\end{lemma}

\section{Proof of Theorem \ref{thm:CGDQuadratic}} \label{app:thm:CGDQuadratic} 

The algorithm in Equation~\eqref{eqn:CGD}  can be written as 
\begin{align*}
x^{k+1} = x^k - \gamma (\nabla f(x^k) + e^k), 
\end{align*}
where $e^k = Q(\nabla f(x^k)) - \nabla f(x^k)$. 
Using that $\nabla f(x^\star) = Hx^\star-b=0$ we have
\begin{align*}
    x^{k+1} - x^\star = (I-\gamma H)(x^k - x^\star) - \gamma e^k,
\end{align*}
or equivalently 
\begin{align}\label{eqn:MainCGD}
    x^k - x^\star = (I-\gamma H)^k(x^0 - x^\star) - \gamma \sum_{i=0}^{k-1} (I-\gamma H)^{k-1-i} e^i.
\end{align}
 By  the triangle inequality and the fact that for a symmetric matrix $I-\gamma H $ we have
 $$\| (I-\gamma H)x \|\leq \rho \| x\| ~~\text{ for all } x\in \mathbb{R}^d.$$
 where 
 $$\rho = \max_{i\in[1,d]} |\lambda_i(I-\gamma H)|= \max_{i\in[1,d]} \vert 1-\gamma \lambda_i\vert $$
 we have
\begin{align*}
    \| x^k - x^\star  \| 
    &\leq \rho^k\|x^0 - x^\star\| + \gamma \sum_{i=0}^{k-1} \rho^{k-1-i}\epsilon.
\end{align*}
%
In particular, when $\gamma = 1/L$ then  $\rho= 1-1/\kappa$ meaning that 
\begin{align*}
\| x^k - x^\star \| &\leq \left(1-{1}/{\kappa}\right)^k\| x^0 - x^\star \| \\ 
&\hspace{0.4cm} + \frac{1}{L} \sum_{i=0}^{k-1} \left(1-{1}/{\kappa}\right)^{k-1-i}\epsilon, 
\end{align*}
where $\kappa = L/\mu$. 
Since  $1-1/\kappa \in (0,1)$ we have
\begin{align*}
 \sum_{i=0}^{k-1} \left(1-{1}/{\kappa}\right)^{k-1-i} & \leq \sum_{i=0}^{\infty} \left(1-{1}/{\kappa}\right)^{i} = \kappa,
\end{align*}
which implies 
\begin{align*}
\| x^k - x^\star \| \leq \left( 1-1/\kappa \right)^k\| x^0 - x^\star \| + \frac{1}{\mu} \epsilon.
\end{align*}
Similarly, when  $\gamma = 2/(\mu +L)$ then $\rho = 1 - 2/{(\kappa+1)}$ and  
\begin{align*}
\| x^k - x^\star \| &\leq \left(1-{2}/{(\kappa+1)}\right)^k\| x^0 - x^\star \| \\ 
&\hspace{0.4cm} + \frac{2}{\mu + L} \sum_{i=0}^{k-1} \left(1-{2}/{(\kappa+1)}\right)^{k-1-i}\epsilon.
\end{align*}
Since $1-2/(\kappa+1) \in (0,1)$ we have  
\begin{align*}
 \sum_{i=0}^{k-1} \left(1-{2}/{(\kappa+1)}\right)^{k-1-i} & \leq \sum_{i=0}^{\infty} \left(1-{2}/{(\kappa+1)}\right)^{i} \\
&= {(\kappa +1)}/{2}.
\end{align*}
 This means that
\begin{align*}
\| x^k - x^\star \| \leq \left( \frac{\kappa-1}{\kappa+1} \right)^k\| x^0 - x^\star \| + \frac{1}{\mu} \epsilon.
\end{align*}

{\color{black}
\section{Proof of Theorem~\ref{thm:3}} \label{App:Pr3}
We can rewrite the direct compression algorithm \eqref{eqn:QDGD} equivalently as  Equation \eqref{eqn:DCSGD_equi} with $\eta^k=0$. By the triangle inequality for the Euclidean norm, 
\begin{align*}
    \| x^{k+1} - x^\star \| \leq \| x^k - x^\star - \gamma \nabla f(x^k) \| + \gamma\frac{1}{n}\sum_{i=1}^n \| e_i^k \|. 
\end{align*}
If $\gamma=2/(\mu+L)$, by the fact that $f(\cdot)$ is $L$-smooth and that $\| e_i^k\| \leq\epsilon$
\begin{align*}
    \| x^{k+1} - x^\star \| \leq \rho \| x^k - x^\star \| + \gamma\epsilon, 
\end{align*}
where $\rho=1-2/(\kappa+1)$ and $\kappa=L/\mu$. By recursively applying this main inequality, we have 
\begin{align*}
    \| x^{k+1} - x^\star \| \leq \rho^k \| x^0 - x^\star \| + \frac{\gamma}{1-\rho}\epsilon. 
\end{align*}
By rearranging the terms, we complete the proof. 
}

{\color{black}
\section{Proof of Theorem~\ref{thm:distCSGD_convex}}\label{app:CGD}
%
%
We can write the algorithm in Equation~\eqref{eqn:QDGD} equivalently as
\begin{align}\label{eqn:DCSGD_equi}
x^{k+1} & = x^k - \gamma \left( \nabla f(x^k) +\eta^k + e^k \right), 
\intertext{where}
\eta^k &=  \frac{1}{n}\sum_{i=1}^n \left[  g_i(x^k;\xi_i^k) - \nabla f_i(x^k)   \right], \quad \text{and} \nonumber \\ 
e^k & = \frac{1}{n}\sum_{i=1}^n   \left[ Q\left( g_i(x^k;\xi_i^k)  \right) -  g_i(x^k;\xi_i^k)  \right]. \nonumber
\end{align}

By Lemma \ref{lemma:norm_sq_trick}, the bounded gradient assumption, and the definition of the $\epsilon$-compressor we have  
\begin{align}
\mathbf{E}\|  \eta^k \|^2 & \leq \frac{1}{n}\sum_{i=1}^n \mathbf{E}\| g_i(x^k;\xi_i^k) - \nabla f_i(x^k) \|^2   \leq \sigma^2, \ \text{and}\label{eqn:etabound} \\
\|  e^k \|^2 & \leq \frac{1}{n}\sum_{i=1}^n \| Q\left( g_i(x^k;\xi_i^k)  \right) -  g_i(x^k;\xi_i^k) \|^2 \leq \epsilon^2.\label{eqn:errorbound}
\end{align}}

 \subsection{Proof of Theorem~\ref{thm:distCSGD_convex}-a)}

By the Lipschitz smoothness assumption on $f(\cdot)$ and Equation~\eqref{eqn:DCSGD_equi} we have
\begin{align*}
f(x^{k+1}) 
&\leq f(x^k) - \gamma \langle \nabla f(x^k), \nabla f(x^k)+\eta^k +e^k  \rangle  \\
&\hspace{0.4cm}+ \frac{L\gamma^2}{2}\|  \nabla f(x^k)+ \eta^k+e^k \|^2. 
\end{align*}
Due to the unbiased property of the stochastic gradient (i.e. $\mathbf{E}  \eta^k=0$), taking the expectation and applying Lemma \ref{lemma:norm_sq_trick}, and Equation~\eqref{eqn:etabound} and \eqref{eqn:errorbound}  yields
%
%
\begin{align*}
\mathbf{E} f(x^{k+1}) 
& \leq  \mathbf{E} f(x^k) - \left(\gamma -\frac{3L\gamma^2}{2}\right)\mathbf{E}\| \nabla f(x^k) \|^2   \\
&\hspace{0.4cm}+ \gamma \mathbf{E}\langle \nabla f(x^k), -e^k  \rangle  +  \frac{3L\gamma^2}{2} ( \sigma^2 + \epsilon^2).
\end{align*}
Next,  applying Lemma \ref{lemma:innerproduct_theta_trick} with $x=\nabla f(x^k)$, $y=-e^k$ and $\theta=1$ into the main result yields 
\begin{align*}
\mathbf{E} f(x^{k+1}) 
&\leq  \mathbf{E} f(x^k) - \left(\frac{\gamma}{2} - \frac{3L\gamma^2}{2} \right)\mathbf{E}\| \nabla f(x^k)\|^2  + T,
\end{align*}
where $T=( 1 + 3L\gamma)\gamma\epsilon^2/2 +3L\gamma^2\sigma^2/2$.
By rearranging  and recalling that $\gamma<1/(3L)$ we get 
\begin{align*}
\mathbf{E}\| \nabla f(x^k)\|^2 
&\leq \frac{2}{\gamma}\frac{1}{1-3L\gamma}\left( \mathbf{E} f(x^k) - \mathbf{E} f(x^{k+1}) + T\right).
\end{align*}
 Using the fact that 
$$\mathop{\min}_{l\in[0,k]} \mathbf{E}\| \nabla f(x^l)\|^2 \leq \frac{1}{k+1}\sum_{l=0}^k \mathbf{E}\| \nabla f(x^l)\|^2$$
 and  the cancelations of telescopic series we get
\begin{align*}
\mathop{\min}\limits_{l\in[0,k]} \mathbf{E}\| \nabla f(x^l)\|^2 
&\leq  \frac{1}{k+1}\frac{2}{\gamma}\frac{1}{1-3L\gamma}\left( \mathbf{E} f(x^0) - \mathbf{E} f(x^{k+1}) \right) \\
&\hspace{0.4cm}+ \frac{2}{\gamma}\frac{1}{1-3L\gamma}T.
\end{align*}
We can now conclude the proof by noting that $ f(x^\star) \leq f(x)$ for all $x\in\mathbb{R}^d$


 \subsection{Proof of Theorem~\ref{thm:distCSGD_convex}-b)}
 
 From the definition of the Euclidean norm and Equation \eqref{eqn:DCSGD_equi}, 
\begin{equation}\label{eqn:Main}
\begin{aligned}
\| x^{k+1} - x^\star \|^2 & = \| x^k - x^\star \|^2 \\  
& \hspace{0.4cm}- 2\gamma \langle \nabla f(x^k) +\eta^k+ e^k, x^k - x^\star \rangle\\
&\hspace{0.4cm}+ \gamma^2 \| \nabla f(x^k) +\eta^k  + e^k\|^2.
\end{aligned}
\end{equation} 
By taking the expected value on both sides and using  the unbiasedness of the stochastic gradient, i.e., that
$$\mathbf{E}\eta^k = \frac{1}{n}\sum_{i=1}^n  \mathbf{E}\left(g_i(x^k;\xi_i^k) - \nabla f_i(x^k)\right)= 0,$$
 and Lemma~\ref{lemma:norm_sq_trick} and Equation~\eqref{eqn:etabound} and~\eqref{eqn:errorbound} to get the bound 
 $$  \| \nabla f(x^k) +\eta^k  + e^k\|^2 \leq 3 \mathbf{E}\| \nabla f(x^k)\|^2  + 3(\sigma^2 +\epsilon^2)$$
  we have 
%
%
%
%
%
%
\begin{align*}
\mathbf{E}\| x^{k+1} - x^\star \|^2 
&\leq \mathbf{E}\| x^k - x^\star \|^2 \\
&\hspace{0.4cm}- 2\gamma \mathbf{E}\langle \nabla f(x^k) + e^k, x^k - x^\star \rangle  \\
& \hspace{0.4cm} + 3\gamma^2 \mathbf{E}\| \nabla f(x^k)\|^2  + 3\gamma^2(\sigma^2 +\epsilon^2).
\end{align*}
 Applying  Equation~\eqref{eqn:mustronglyconvex} with $x=x^k$ and $y= x^\star$ and using Lemma \ref{lemma:Nesterov_trick} with $x=x^k$ we have
%
%
%
%
%
%
\begin{align*}
\mathbf{E}\| x^{k+1} - x^\star \|^2 
&\leq (1-\mu\gamma)\mathbf{E}\| x^k - x^\star \|^2 \\
&\hspace{0.4cm}- 2(\gamma  - 3L\gamma^2)\mathbf{E}[f(x^k) - f(x^\star)] \\ 
& \hspace{0.4cm} + 2\gamma \mathbf{E}\langle  e^k, x^\star - x^k \rangle  + 3\gamma^2(\sigma^2 +\epsilon^2).
\end{align*}
From Lemma~\ref{lemma:innerproduct_theta_trick} with $\theta=\mu$ and Equation~\eqref{eqn:errorbound}, we have
\begin{align*}
2\gamma \langle e^k , x^\star - x^k \rangle \leq \mu\gamma \|  x^k - x^\star \|^2 + \epsilon^2 \gamma/\mu, 
\end{align*}
which yields 
\begin{align*}
\mathbf{E}\| x^{k+1} - x^\star \|^2 &\leq \mathbf{E}\| x^k - x^\star \|^2 \\
&\hspace{0.4cm}- 2\gamma(1  - 3L\gamma)\mathbf{E}[f(x^k) - f(x^\star)]  + T,
\end{align*}
where $T= \gamma(1/\mu + 3\gamma)\epsilon^2  + 3\gamma^2\sigma^2$.
 By  rearranging the terms and recalling that $\gamma<1/3L$ we get
\begin{align*}
&\mathbf{E}\left(  f(x^k) - f(x^\star) \right)  \\
&\hspace{0.4cm} \leq \frac{1}{2\gamma}\frac{1}{1-3L\gamma}\left( \mathbf{E}\| x^k - x^\star \|^2 - \mathbf{E}\| x^{k+1} - x^\star \|^2  + T\right).
\end{align*}
Define $\bar x^k = \sum_{l=0}^k x^l/(k+1)$. By the convexity of $f(\cdot)$ and from the cancelations of the telescopic series we have
\begin{align*}
\mathbf{E}\left(  f(\bar x^k) - f(x^\star) \right)  
& \leq \frac{1}{k+1}\sum_{l=0}^k \mathbf{E}\left(  f(x^l) - f(x^\star) \right)  \\
& \leq \frac{1}{k+1} \frac{1}{2\gamma}\frac{1}{1-3L\gamma}\| x^0 - x^\star\|^2\\
&\hspace{0.4cm}+ \frac{1}{2\gamma}\frac{1}{1-3L\gamma}T.
\end{align*}
Hence, the proof is complete.

\section{Proof of Theorem \ref{thm:ErrorCompensatedSC}}\label{app:thm:ErrorCompensatedSC}

We can write the algorithm in Equation~\eqref{eqn:ECCGD} equivalently as
\begin{align*}
x^{k+1} &= x^k - \gamma (\nabla f(x^k) - c^k), \intertext{where}
c^k & =  \nabla f(x^k) - Q(\nabla f(x^k) + A_\gamma e^k) , \quad \text{and} \\ 
e^{k+1} & = c^k + A_\gamma e^k
\end{align*}
and $A_\gamma=I - \gamma H$. Following similar line of arguments as in the proof of Theorem~\ref{thm:CGDQuadratic} we obtain
\begin{align*}
    x^k - x^\star = A_\gamma^k(x^0 - x^\star) + \gamma \sum_{i=0}^{k-1} A_\gamma^{k-1-i} c^i.
\end{align*}
By using that $e^k = \sum_{i=0}^{k-1} A_\gamma^{k-1-i}c^i$ and $e^0=0$ we get that
\begin{align*}
     x^k - x^\star = A_\gamma^k(x^0 - x^\star) + \gamma e^k.
\end{align*}
Since $A_\gamma$ is symmetric, by  the triangle inequality and the fact that $\| e^k\|\leq\epsilon$ (since $e^k$ is the compression error)  we have
\begin{align*}
    \| x^k - x^\star \| \leq \rho^k \| x^0 -x^\star \| + \gamma\epsilon,
\end{align*}
where $\rho=\max_{i\in[1,d]}\vert 1-\gamma\lambda_i \vert$. 
 Now following similar arguments as used in the proof of Theorem~\ref{thm:CGDQuadratic} 
  If $\gamma  =1/L$ then $\rho = 1-1/\kappa$. Since $1-1/\kappa \in (0,1)$  
  we have
\begin{align*}
    \| x^k - x^\star \| \leq \left( 1- \frac{1}{\kappa} \right)^k \| x^0 -x^\star \| + \frac{1}{L}\epsilon.
\end{align*}
If $\gamma = 2/(\mu+L)$ then $\rho = 1-2/(\kappa+1)$. Since $1-2/(\kappa+1)\in(0,1)$ we have
\begin{align*}
    \| x^k - x^\star \| \leq \left( \frac{\kappa-1}{\kappa+1} \right)^k \| x^0 -x^\star \| + \frac{2}{\mu+L}\epsilon.
\end{align*}

{\color{black}
\section{Proof of Theorem~\ref{thm:5}} \label{App:Pr5}

We can rewrite the error compensation algorithm \eqref{eqn:DECCSGD} with $g_i(x^k;\xi_i^k) = \nabla f_i(x^k)$ and $A_i^k = I - \gamma \nabla^2 f_i(x^k)$ equivalently as Equation \eqref{eqn:ECCSGD_equi} with $\eta^k=0$. By the triangle inequality for the Euclidean norm,
\begin{align*}
    \| \tilde x^{k+1} - x^\star \| 
    &\leq  \| \tilde x^k - x^\star - \gamma \nabla f(\tilde x^k) \| \\
    &\hspace{0.4cm}+ \gamma \| \nabla f(\tilde x^k)  - \nabla f(x^k) \| \\
    &\hspace{0.4cm}+ \gamma^2 \frac{1}{n}\sum_{i=1}^n \| \nabla^2 f_i(x^k) e_i^k \|.
\end{align*}
If $\gamma=2/(\mu+L)$, by the fact that $f(\cdot)$ is $L$-smooth and that $\tilde x^k - x^k = -\gamma \sum_{i=1}^n e_i^k/n$
\begin{align*}
    \| \tilde x^{k+1} - x^\star \| 
    &\leq  \rho \| \tilde x^k - x^\star \| + \gamma^2 L \epsilon \\
    &\hspace{0.4cm}+ \gamma^2 \frac{1}{n}\sum_{i=1}^n \| \nabla^2 f_i(x^k) e_i^k \|,
\end{align*}
where $\rho=1-2/(\kappa+1)$ and $\kappa=L/\mu$.
Since each $f_i(\cdot)$ is $L$-smooth, $\nabla^2 f_i(x)  \preceq L I$ for $x\in\mathbb{R}^d$, and $\| e_i^k\|\leq\epsilon$, we have 
\begin{align*}
    \| \tilde x^{k+1} - x^\star \| 
    &\leq  \rho \| \tilde x^k - x^\star \| + 2\gamma^2 L \epsilon.
\end{align*}
By recursively applying this main inequality, 
\begin{align*}
    \| \tilde x^{k} - x^\star \| 
    &\leq  \rho^k \| \tilde x^0 - x^\star \| + \frac{2\gamma^2 L}{1-\rho} \epsilon.
\end{align*}
Using the triangle inequality and the fact that $\| x^k-\tilde x^k \|\leq\gamma\epsilon$, we can conclude that 
\begin{align*}
    \|x^k  - x^\star\| 
    &\leq \| \tilde x^k -x^\star\| + \| x^k-\tilde x^k \| \\
    & \leq \rho^k \| \tilde x^0 - x^\star \| + \left(\frac{2\gamma^2 L}{1-\rho} + \gamma \right)\epsilon.
\end{align*}
Since $\tilde x^0=x^0$, the proof is complete.}

\section{Proof of Theorem~\ref{thm:distEC-CSGD_convex}} \label{app:ECCGD}



 We can write the algorithm in Equation~\eqref{eqn:DECCSGD} equivalently as
\begin{align}\label{eqn:ECCSGD_equi}
\tilde x^{k+1} &= \tilde x^k - \gamma \left[ \nabla f(x^k) +\eta^k \right] - \gamma \frac{1}{n}\sum_{i=1}^n (A_i^k - I )e_i^k,
\intertext{where}
\tilde x^k &= x^k - \gamma\frac{1}{n}\sum_{i=1}^n e_i^k, \quad \text{and} \nonumber \\ 
\eta^k & =  \frac{1}{n}\sum_{i=1}^n \left[ \nabla g_i(x^k;\xi_i^k) - \nabla f_i(x^k) \right]. \nonumber
\end{align}	
By Lemma \ref{lemma:norm_sq_trick},  the bounded gradient assumption and by the definition of the $\epsilon$-compressor, it  can be proved that
\begin{align}
\mathbf{E}\|  \eta^k \|^2 &  \leq \sigma^2, \quad\text{and}\label{eqn:etaboundEC}\\
\left\|  x^k - \tilde x^k \right\|^2  & \leq \gamma^2\sum_{i=1}^n \| e_i^k \|^2/n \leq \gamma^2\epsilon^2. \label{eqn:errorboundEC}
\end{align}
%
 

\subsection{Proof of Theorem~\ref{thm:distEC-CSGD_convex}-a)}
Before deriving the main result we prove two lemmas that are need in our analysis.
%
\begin{lemma}\label{lemma:trick_ECCSGDHessian}
Assume that $\|e_i^k\|\leq \epsilon$, and that the Hessian $H_i^k$ satisfies the unbiased and bounded variance assumptions described in Equation~\eqref{eqn:unbiasedHessian}  and~\eqref{eqn:boundedVRHessian}. If $\nabla^2 f_i(x) \preccurlyeq L I$ for $x\in\mathbb{R}^d$, then
\begin{align}\label{eqn:trick_ECCSGDHessian2}
\mathbf{E}\left\| \gamma \frac{1}{n}\sum_{i=1}^n H_i^k e_i^k \right\|^2 \leq 2\gamma^2(\sigma_H^2 + L^2)\epsilon^2 , \quad \text{for } k\in\mathbb{N}.
\end{align}
\end{lemma}
\begin{proof}
By Lemma  \ref{lemma:norm_sq_trick}, we have
\begin{equation*}
\begin{aligned}
\mathbf{E}\left\| \gamma \frac{1}{n}\sum_{i=1}^n H_i^k e_i^k \right\|^2&\leq 2\gamma^2 \frac{1}{n} \sum_{i=1}^n \mathbf{E}\| [H_i^k-\nabla^2 f_i(x^k) ]e_i^k \|^2  \\
&\hspace{0.4cm} + 2\gamma^2 \frac{1}{n} \sum_{i=1}^n \mathbf{E}\| \nabla^2 f_i(x^k) e_i^k \|^2.
\end{aligned}
\end{equation*}
Since $H_i^k -\nabla^2 f_i(x^k)$ is symmetric, using Equation~\eqref{eqn:boundedVRHessian}, and the fact that $\nabla^2 f_i(x^k) \preccurlyeq L I$ and that $\|e_i^k\|\leq \epsilon$ yields
\begin{equation*}
\begin{aligned}
&\mathbf{E}\left\| \gamma \frac{1}{n}\sum_{i=1}^n H_i^k e_i^k \right\|^2 \leq 2\gamma^2 (\sigma_H^2 +L^2)\epsilon^2. 
\end{aligned}
\end{equation*}
\end{proof}
%
\begin{lemma}\label{lemma:stronglyConvexTrick_Main}
If $f(\cdot)$ is strongly convex, then for $\theta_1>0$
\begin{align}\label{eqn:stronglyConvexTrick_Main}
 - \langle \nabla f(x^k), \tilde x^{k} - x^\star \rangle  & \leq - (f(x^k) - f(x^\star)) - \frac{\mu}{4}\| \tilde x^k - x^\star \|^2 \notag\\
 &\hspace{0.4cm} + \frac{1}{2}\left(\mu+\frac{1}{\theta_1}\right)\| \tilde x^k - x^k \|^2   \notag \\
 &\hspace{0.4cm} + \frac{\theta_1}{2}\| \nabla f(x^k) \|^2.
\end{align}
\end{lemma}
\begin{proof}
By using the strong convexity inequality in Equation~\eqref{eqn:mustronglyconvex} with $x=x^k$ and $y= x^\star$ we have
\begin{align*}
 - \langle \nabla f(x^k), x^{k} - x^\star \rangle \leq - (f(x^k) - f(x^\star)) - \frac{\mu}{2}\| x^k - x^\star \|^2.
\end{align*}
Using the fact that $\|  x+ y\|^2\leq 2\|x\|^2 + 2\| y\|^2$ with $x=x^k-x^\star$ and $y=\tilde x^k - x^k$, we have
\begin{align*}
 - \| x^k - x^\star \|^2 \leq  - \frac{1}{2}\| \tilde x^k - x^\star \|^2 +  \| x^k - \tilde x^k \|^2 .
\end{align*}
Combining these inequalities yields
\begin{equation} \label{eqn:stronglyConvexTrick_Ri}
\begin{aligned}
 - \langle \nabla f(x^k), x^{k} - x^\star \rangle 
 &\leq - (f(x^k) - f(x^\star))  \\ 
 &\hspace{0.4cm} - \frac{\mu}{4}\| \tilde x^k - x^\star \|^2 + \frac{\mu}{2}\| x^k - \tilde x^k \|^2. 
\end{aligned}
\end{equation}

Next,  by Lemma \ref{lemma:innerproduct_theta_trick} 
\begin{align}\label{eqn:stronglyConvexTrick_Ri2}
 -\langle \nabla f(x^k),   \tilde x^k - x^{k} \rangle \leq  \frac{1}{2\theta_1}\| x^{k} - \tilde x^k \|^2 +\frac{\theta_1}{2} \| \nabla f(x^k)\|^2,
\end{align}
for $\theta_1>0$. Summing Equation~\eqref{eqn:stronglyConvexTrick_Ri} and~\eqref{eqn:stronglyConvexTrick_Ri2} completes the proof.
\end{proof}

By using the $L$-smoothness of $f(\cdot)$ and Equation~\eqref{eqn:ECCSGD_equi} with $A_i^k$ defined by Equation~\eqref{eqn:defOfAik} we have
\begin{align*}
f(\tilde x^{k+1}) 
&\leq f(\tilde x^{k}) - \gamma \langle \nabla f(\tilde x^k), \nabla f(x^k)+\eta^k \rangle \\
&\hspace{0.4cm}+ \gamma \left\langle \nabla f(\tilde x^k), \gamma \frac{1}{n} \sum_{i=1}^n H^k_i e^k_i \right\rangle  \\
& \hspace{0.4cm}+ \frac{L\gamma^2}{2}\left\| \nabla f(x^k) +\eta^k - \gamma \frac{1}{n}\sum_{i=1}^n H^k_i e^k_i \right \|^2.
\end{align*}

By the unbiased property of the stochastic gradient in Equation~\eqref{eqn:unbiasednessSGD}, and by applying Lemma~\ref{lemma:innerproduct_theta_trick}  with $\theta=1$ and Lemma~\ref{lemma:norm_sq_trick}   we get
\begin{align*}
\mathbf{E}f(\tilde x^{k+1}) 
&\leq \mathbf{E} f(\tilde x^{k}) - \gamma \mathbf{E}\langle \nabla f(\tilde x^k), \nabla f(x^k)\rangle \\
&\hspace{0.4cm} + \left( \frac{\gamma}{2}  + \frac{3L\gamma^2}{2}\right)\mathbf{E}\| \nabla f(\tilde x^k) \|^2 + \frac{3L\gamma^2}{2}\mathbf{E}\| \eta^k \|^2 \\
& \hspace{0.4cm}+ \left( \frac{\gamma}{2} +  \frac{3L\gamma^2}{2} \right)\mathbf{E}\left\| \gamma \frac{1}{n}\sum_{i=1}^n H^k_i e^k_i \right \|^2.
\end{align*}
Since each $f_i(\cdot)$ is $L$-smooth, $\nabla^2 f_i(x)  \preceq L I$ for $x\in\mathbb{R}^d$. 
Applying the bounds in Equation~\eqref{eqn:etaboundEC} and~\eqref{eqn:trick_ECCSGDHessian2}  yields 
\begin{align*}
\mathbf{E}f(\tilde x^{k+1}) 
&\leq \mathbf{E}f(\tilde x^{k}) - \gamma\mathbf{E} \langle \nabla f(\tilde x^k), \nabla f(x^k) \rangle \\
&\hspace{0.4cm}+\left( \frac{\gamma}{2}  +\frac{3L\gamma^2}{2} \right)\mathbf{E}\| \nabla f(x^k)\|^2 + T, 
\end{align*}
where $T={3L\gamma^2}\sigma^2/2  + (1+3L\gamma)(\sigma_H^2+L^2)\gamma^3\epsilon^2.$
 Using  that  $$-2\langle x,y\rangle = -\| x \|^2-\| y\|^2+\| x-y \|^2 \quad \text{for all } x,y\in\mathbb{R}^d$$
  we have
\begin{align*}
\mathbf{E}f(\tilde x^{k+1}) 
&\leq \mathbf{E}f(\tilde x^{k})  -\left( \frac{\gamma}{2} - \frac{3L\gamma^2}{2} \right) \mathbf{E}\| \nabla f( x^k) \|^2 \\ 
& \hspace{0.4cm} + \frac{\gamma}{2}\mathbf{E}\| \nabla f(\tilde x^k)- \nabla f( x^k)  \|^2  + T.
\end{align*} 
By the Lipschitz continuity assumption of $\nabla f(\cdot)$,  and by \eqref{eqn:errorboundEC}, 
\begin{align*}
\mathbf{E}f(\tilde x^{k+1}) 
&\leq \mathbf{E}f(\tilde x^{k})  -\left( \frac{\gamma}{2} - \frac{3L\gamma^2}{2} \right) \mathbf{E}\| \nabla f( x^k) \|^2  + \bar T,
\end{align*} 
where $\bar T = T+ L^2(\gamma^3/2)\epsilon^2$.
By rearranging the terms and recalling that $\gamma < 1/(3L)$  we get
\begin{align*}
\mathbf{E}\| \nabla f( x^k) \|^2 
& \leq \frac{2}{\gamma}\frac{1}{1-3L\gamma}\left(\mathbf{E}f(\tilde x^{k}) - \mathbf{E}f(\tilde x^{k+1}) + \bar T \right).
\end{align*}
Since $\mathop{\min}_{l\in[0,k]} \mathbf{E}\| \nabla f(x^l)\|^2 \leq \sum_{l=0}^k \mathbf{E}\| \nabla f(x^l)\|^2/(k+1)$, we have
\begin{align*}
\mathop{\min}\limits_{l\in[0,k]} \mathbf{E} \| \nabla f(x^l) \|^2 
& \leq \frac{1}{k+1}\frac{2}{\gamma}\frac{1}{1-3L\gamma}\left(\mathbf{E}f(\tilde x^{0}) - \mathbf{E}f(\tilde x^{k+1}) \right) \\
&\hspace{0.4cm} + \frac{2}{\gamma}\frac{1}{1-3L\gamma}\bar T.
\end{align*}
By the fact that $e^0=0$ (i.e. $\tilde x^0 = x^0$), that $f(x)\geq f(x^\star)$ for $x\in\mathbb{R}^d$ we complete the proof.


\subsection{Proof of Theorem~\ref{thm:distEC-CSGD_convex}-b)}

From Equation~\eqref{eqn:ECCSGD_equi}  with $A_i^k$ defined by Equation~\eqref{eqn:defOfAik}  we have
\begin{align*}
& \| \tilde x^{k+1} - x^\star \|^2 \\
&\hspace{0.4cm} = \| \tilde x^{k} - x^\star \|^2 - 2\gamma \langle \nabla f(x^k) + \eta^k, \tilde x^{k} - x^\star \rangle \\
&\hspace{0.8cm} + 2\gamma \left\langle \gamma\frac{1}{n} \sum_{i=1}^n H_i^k e^k_i, \tilde x^{k} - x^\star \right \rangle  \\
&\hspace{0.8cm} + \gamma^2\left\| \nabla f(x^k) +\eta^k -  \gamma\frac{1}{n} \sum_{i=1}^n H_i^k e^k_i \right\|^2.
\end{align*}
By the unbiasedness of the stochastic gradient described in Equation~\eqref{eqn:unbiasednessSGD}, by Lemma~\ref{lemma:norm_sq_trick}, by Lemma~\ref{lemma:innerproduct_theta_trick} with $\theta=\mu/2$ and by the bound in Equation~\eqref{eqn:etaboundEC} we have
\begin{align*}
&\mathbf{E}\| \tilde x^{k+1} - x^\star \|^2 \\
&\hspace{0.4cm} \leq \left(1+\frac{\mu\gamma}{2}\right)\mathbf{E}\| \tilde x^{k} - x^\star \|^2 - 2\gamma\mathbf{E} \langle \nabla f(x^k), \tilde x^{k} - x^\star \rangle \\
&\hspace{0.8cm} + \left(\frac{2\gamma}{\mu} + 3\gamma^2\right)\mathbf{E}\left\| \gamma\frac{1}{n} \sum_{i=1}^n H_i^k e^k_i \right\|^2\\
&\hspace{0.8cm}+ 3\gamma^2\mathbf{E}\| \nabla f(x^k) \|^2 + 3\gamma^2 \sigma^2.
\end{align*}
%
%
%
%
Since each $f_i(\cdot)$ is $L$-smooth, $\nabla^2 f_i(x)  \preceq L I$ for $x\in\mathbb{R}^d$ so we can apply Lemma~ 
\ref{lemma:stronglyConvexTrick_Main}. 
 From Equation~\eqref{eqn:errorboundEC} in Lemma~\ref{lemma:trick_ECCSGDHessian} and Equation~\eqref{eqn:stronglyConvexTrick_Main} in Lemma~\ref{lemma:stronglyConvexTrick_Main} with $\theta_1=\beta/L$ we have 
\begin{align*}
& \mathbf{E}\| \tilde x^{k+1} - x^\star \|^2 \\
&\hspace{0.4cm} \leq \mathbf{E}\| \tilde x^{k} - x^\star \|^2 - 2\gamma\mathbf{E}\left(f(x^k) - f(x^\star) \right) \\
&\hspace{0.8cm}+ \left( \frac{\beta \gamma}{L}+3\gamma^2\right)\mathbf{E}\| \nabla f(x^k) \|^2 + \bar  T, \intertext{where}
& \bar T  = \left( \mu+\frac{L}{\beta} + \left(\frac{4}{\mu} + 6\gamma\right)(\sigma_H^2+L^2) \right)\gamma^3\epsilon^2+ 3\gamma^2\sigma^2.
\end{align*}
By Lemma \ref{lemma:Nesterov_trick}, we have
\begin{align*}
&\mathbf{E}\| \tilde x^{k+1} - x^\star \|^2 \\ 
&\hspace{0.4cm}\leq \mathbf{E}\| \tilde x^{k} - x^\star \|^2 - 2\alpha\gamma\mathbf{E}\left(f(x^k) - f(x^\star) \right) + \bar T.
\end{align*}
where $\alpha = 1 -  \beta - 3L \gamma$. By recalling that $\gamma<(1-\beta)/(3L)$ and $\beta\in (0,1)$ then 
\begin{align*}
& \mathbf{E}\left(f(x^k) - f(x^\star) \right) \\
& \hspace{0.4cm} \leq \frac{1}{2\alpha\gamma} \left(\mathbf{E}\| \tilde x^{k} - x^\star \|^2 - \mathbf{E}\| \tilde x^{k+1} - x^\star \|^2 + \bar T \right).
\end{align*}
Define $\bar x^k = \sum_{l=0}^k x^l/(k+1)$. By the convexity of $f(\cdot)$ and the cancelations in  the telescopic series  we have
\begin{align*}
\mathbf{E}\left(f(\bar x^k) - f(x^\star) \right) 
 & \leq \frac{1}{k+1}\sum_{l=0}^k \mathbf{E}\left(f( x^l) - f(x^\star) \right) \\ 
 &\leq \frac{1}{k+1}\frac{1}{2\alpha\gamma} \mathbf{E}\| \tilde x^{0} - x^\star \|^2 + \frac{1}{2\alpha\gamma} \bar T.
\end{align*}
By the fact that $e^0=0$ (i.e. $\tilde x^0 = x^0$), the proof is complete. 

\end{document}


\appendix
\section{Review of Useful Lemmas}
This section states lemmas which establish convergence rate results. 

\begin{lemma}\label{lemma:Convex_C}
Consider the optimization problem \eqref{eqn:Problem} under Assumption \ref{assum:fiConvex}. Assume that $g_k:\mathbb{R}^d\rightarrow\mathbb{R}^d$ satisfies  
\begin{align*}
\mathbf{E}\{g_k \}= \nabla f(x_k)  \quad \text { and  } \quad \mathbf{E} \| g_k\|^2 \leq\alpha_1\cdot\mathbf{E}\langle \nabla f(x_k) - \nabla f(x^\star), x_k - x^\star  \rangle + \alpha_2,
\end{align*} 
for positive constants $\alpha_1,\alpha_2$. Then, the iterates $\{x_k\}_{k\in\mathbb{N}}$ generated by the recursion \[x_{k+1} = x_k - \gamma g_k\] 
with $\gamma < 2/\alpha_1$ satisfy
\begin{align*}
\mathbf{E} \{  f(\bar x_k) - f(x^\star)\} \leq \frac{1}{k+1}\cdot\frac{1}{\gamma}\cdot\frac{1}{2-\alpha_1\gamma} \|  x_0 - x^\star\|^2 +\frac{\gamma}{2-\alpha_1\gamma}  \alpha_2,
\end{align*}
where $\bar x_k = (1/(k+1))\sum_{l=0}^k x_l$.
\end{lemma}
\begin{proof}
From the definiton of the Euclidean norm and the recursion, 
\begin{align*}
\| x_{k+1} - x^\star \|^2 & = \|  x_k - x^\star\|^2 - 2\gamma \left\langle g_k, x_k -x^\star \right\rangle + \gamma^2 \left\| g_k \right\|^2.
\end{align*}
Taking the expectation with respect to all the randomness in the algorithm yields 
\begin{align*}
\mathbf{E}\| x_{k+1} - x^\star \|^2 & = \mathbf{E}\|  x_k - x^\star\|^2 - 2\gamma \mathbf{E}\left\langle \nabla f(x_k), x_k -x^\star \right\rangle + \gamma^2 \mathbf{E}\left\| g_k \right\|^2.
\end{align*}
Using the second property of $g_k$ and the fact that $\nabla f(x^\star)=0$ yield
\begin{align*}
\mathbf{E}\| x_{k+1} - x^\star \|^2 & \leq \mathbf{E}\|  x_k - x^\star\|^2 - (2\gamma -\alpha_1 \gamma^2)\mathbf{E}\left\langle \nabla f(x_k), x_k -x^\star \right\rangle + \gamma^2 \alpha_2.
\end{align*} 
Assume that $\gamma < 2/\alpha_1$. By the convexity of $f$, we obtain
\begin{align*}
\mathbf{E} \{  f(x_k) - f(x^\star)\} \leq \frac{1}{\gamma}\frac{1}{2-\alpha_1\gamma} \left( \mathbf{E}\|  x_k - x^\star\|^2 -  \mathbf{E}\|  x_{k+1} - x^\star\|^2 \right) +\frac{\gamma}{2-\alpha_1L\gamma}  \alpha_2.
\end{align*}
Define $\bar x_k = (1/(k+1))\sum_{l=0}^k x_l$. Again by the convexity of $f$, we have
\begin{align*}
\mathbf{E} \{  f(\bar x_k) - f(x^\star)\} 
& \leq \frac{1}{k+1} \sum_{l=0}^k\mathbf{E} \{  f( x_l) - f(x^\star)\} \\  
&\leq \frac{1}{k+1}\cdot\frac{1}{\gamma}\cdot\frac{1}{2-\alpha_1\gamma} \left( \mathbf{E}\|  x_0 - x^\star\|^2 -  \mathbf{E}\|  x_{k+1} - x^\star\|^2 \right) +\frac{\gamma}{2-\alpha_1\gamma}  \alpha_2.
\end{align*}
Since $\|x\|^2\geq 0$ for $x\in\mathbb{R}^d$, we reach the result. 
\end{proof}

\begin{lemma}\label{lemma:NonConvex_C}
Consider the optimization problem \eqref{eqn:Problem} where $f$ is $L-$smooth. Assume that $g_k:\mathbb{R}^d\rightarrow\mathbb{R}^d$ satisfies  
\begin{align*}
\mathbf{E}\{g_k \}= \nabla f(x_k)  \quad \text { and  } \quad \mathbf{E} \| g_k\|^2 \leq \alpha_1 \mathbf{E}\| \nabla f(x_k) \|^2 + \alpha_2,
\end{align*} 
for positive constants $\alpha_1,\alpha_2$. Then, the iterates $\{x_k\}_{k\in\mathbb{N}}$ generated by the recursion \[x_{k+1} = x_k - \gamma g_k\] 
with $\gamma < 2/(\alpha_1 L)$ satisfy
\begin{align*}
\min_{l\in\{0,1,\ldots,k\}}\mathbf{E} \| \nabla f(x_l) \|^2 
\leq \frac{1}{k+1}\cdot\frac{1}{\gamma} \frac{1}{1-\alpha_1L/2}\left( f(x_0) - f(x^\star)\right) + \frac{L\gamma}{1-\alpha_1L/2}\alpha_2.
\end{align*}
\end{lemma}
\begin{proof}
By the smoothness assumption of $f$ and the recursion, 
\begin{align*}
f(x_{k+1}) \leq f(x_k) - \gamma \left\langle \nabla f(x_k), g_k    \right\rangle + \frac{L\gamma^2}{2} \left\| g_k \right\|^2.
\end{align*}
Taking the expectation with respect to all the randomness in the algorithm yields 
\begin{align*}
\mathbf{E} f(x_{k+1}) \leq \mathbf{E} f(x_k) - \gamma \mathbf{E} \| \nabla f(x_k) \|^2 + \frac{L\gamma^2}{2}\mathbf{E}\| g_k\|^2.
\end{align*}
By the second property of $g_k$, we have 
\begin{align*}
\mathbf{E} f(x_{k+1}) \leq \mathbf{E} f(x_k) - \left(\gamma -\alpha_1 L\gamma^2/2 \right)\mathbf{E} \| \nabla f(x_k) \|^2 + \frac{L\gamma^2}{2}\alpha_2.
\end{align*}
Assume that $\gamma < 2/(\alpha_1 L)$. Then, 
\begin{align*}
\mathbf{E} \| \nabla f(x_k) \|^2 \leq \frac{1}{\gamma} \frac{1}{1-\alpha_1L/2}\left( \mathbf{E} f(x_k) - \mathbf{E} f(x_{k+1})\right) + \frac{L\gamma}{1-\alpha_1L/2}\alpha_2,
\end{align*}
or equivalently 
\begin{align*}
\min_{l\in\{0,1,\ldots,k\}}\mathbf{E} \| \nabla f(x_l) \|^2 
& \leq \frac{1}{k+1} \sum_{l=0}^k \mathbf{E} \| \nabla f(x_l) \|^2 \\
&\leq \frac{1}{k+1}\cdot\frac{1}{\gamma} \frac{1}{1-\alpha_1L/2}\left( \mathbf{E} f(x_0) - \mathbf{E} f(x_{k+1})\right) + \frac{L\gamma}{1-\alpha_1L/2}\alpha_2.
\end{align*}
Since $f(x)\geq f(x^\star)$ for $x\in\mathbb{R}^d$, the proof is complete. 
\end{proof}

\begin{lemma}\label{lemma:Convex_EC}
Consider the optimization problem \eqref{eqn:Problem} where $f$ is $L-$smooth and Assumption \ref{assum:fiConvex} holds. Assume that $g_k:\mathbb{R}^d\rightarrow\mathbb{R}^d$ satisfies  
\begin{align*}
\mathbf{E}\{g_k \}= \nabla f(x_k)  \quad \text { and  } \quad \mathbf{E} \| g_k\|^2 \leq\alpha_1\cdot\mathbf{E}\langle \nabla f(x_k) - \nabla f(x^\star), x_k - x^\star  \rangle + \alpha_2,
\end{align*} 
for positive constants $\alpha_1,\alpha_2$. Then, the iterates $\{x_k\}_{k\in\mathbb{N}}$ generated by the recursion \[\tilde x_{k+1} = \tilde x_k - \gamma g_k\] 
where $\tilde x_0 = x_0$, $\mathbf{E}\| x_k -\tilde x_k\|^2\leq\beta$ for a positive constant $\beta$, and  $\gamma <1/\alpha_1$ satisfy
\begin{align*}
\mathbf{E} \{  f(\bar x_k) - f(x^\star)\} 
&\leq \frac{1}{k+1}\frac{1}{\gamma} \frac{1}{1-\alpha_1\gamma}\|  x_0  -x^\star \|^2  + \frac{2L}{1-\alpha_1\gamma}\beta + \frac{\gamma}{1-\alpha_1\gamma}\alpha_2.
\end{align*}
where $\bar x_k = (1/(k+1))\sum_{l=0}^k x_l$.
\end{lemma}
\begin{proof}
From the definiton of the Euclidean norm and the recursion, 
\begin{align*}
\| \tilde x_{k+1} - x^\star \|^2 &= \| \tilde x_k  -x^\star \|^2 - 2\gamma\left\langle g_k, \tilde x_k - x^\star \right\rangle + \gamma^2 \left\| g_k \right\|^2.
\end{align*}
Taking the expectation over all the randomness in the algorithm yields 
\begin{align*}
\mathbf{E}\| \tilde x_{k+1} - x^\star \|^2 = \mathbf{E}\| \tilde x_k  -x^\star \|^2 - 2\gamma\mathbf{E}\left\langle \nabla f(x_k), \tilde x_k - x^\star \right\rangle + \gamma^2 \mathbf{E}\left\|g_k \right\|^2.
\end{align*}
Using the second inequality of $g_k$ and the fact that $\nabla f(x^\star)=0$ yield  
\begin{align*}
\mathbf{E}\| \tilde x_{k+1} - x^\star \|^2 
&\leq \mathbf{E}\| \tilde x_k  -x^\star \|^2 - 2\gamma\mathbf{E}\left\langle \nabla f(x_k), \tilde x_k - x^\star \right\rangle  \\
&\hspace{0.4cm}+ \alpha_1\gamma^2 \mathbf{E}\left\langle \nabla f(x_k), x_k - x^\star \right\rangle + \gamma^2\alpha_2,
\end{align*}
or equivalently 
\begin{align*}
\mathbf{E}\| \tilde x_{k+1} - x^\star \|^2 
&\leq \mathbf{E}\| \tilde x_k  -x^\star \|^2 - (2\gamma-\alpha_1\gamma^2)\mathbf{E}\left\langle \nabla f(x_k), x_k - x^\star \right\rangle  \\
&\hspace{0.4cm} -2\gamma \mathbf{E}\left\langle \nabla f(x_k), \tilde x_k - x_k \right\rangle + \gamma^2\alpha_2.
\end{align*}
By the convexity of the objective function $f$ and by the fact that $-2\langle x,y\rangle \leq \theta\| x \|^2 +(1/\theta)\|y\|^2$ for $x,y\in\mathbb{R}^d$ and $\theta>0$, 
\begin{align*}
    -\langle \nabla f( x_k),  x_k -x^\star \rangle & \leq -[f(x_k) - f(x^\star)], \quad \text{and} \\
    -\langle \nabla f(x_k),\tilde x_k -  x_k \rangle & \leq  \frac{\theta}{2}\| \tilde x_k - x_k \|^2 + \frac{1}{2\theta}\| \nabla f(x_k) \|^2,  \\ 
\end{align*}
respectively. Therefore, 
\begin{align*}
\mathbf{E}\| \tilde x_{k+1} - x^\star \|^2 
&\leq \mathbf{E}\| \tilde x_k  -x^\star \|^2 - (2\gamma-\alpha_1\gamma^2)\mathbf{E} \{f(x_k) - f(x^\star) \}\\
&\hspace{0.4cm} + \frac{\gamma}{\theta} \mathbf{E}\| \nabla f(x_k) \|^2 + \gamma\theta \mathbf{E}\| x_k -\tilde x_k \|^2+ \gamma^2\alpha_2.
\end{align*}
By the coercitivity of $f$, i.e. $\| \nabla f(x) \|^2 \leq 2L[ f(x) - f(x^\star)]$ for $x\in\mathbb{R}^d$, we have 
\begin{align*}
\mathbf{E}\| \tilde x_{k+1} - x^\star \|^2 
&\leq \mathbf{E}\| \tilde x_k  -x^\star \|^2 - (2\gamma-2L\gamma/\theta-\alpha_1\gamma^2)\mathbf{E} \{f(x_k) - f(x^\star) \}\\
&\hspace{0.4cm} + \gamma\theta \mathbf{E}\| x_k -\tilde x_k \|^2+ \gamma^2\alpha_2.
\end{align*}
Assume that $\mathbf{E}\|  x_k - \tilde x_k \|^2 \leq \beta$ for a positive constant $\beta$, and $\theta = 2L$ and then $\gamma < 1/\alpha_1$. Then, we have
\begin{align*}
\mathbf{E} \{f(x_k) - f(x^\star) \} 
&\leq \frac{1}{\gamma} \frac{1}{1-\alpha_1\gamma}\left( \mathbf{E}\| \tilde x_k  -x^\star \|^2 - \mathbf{E}\| \tilde x_{k+1}  -x^\star \|^2 \right) \\
& \hspace{0.4cm}+ \frac{2L}{1-\alpha_1\gamma}\beta + \frac{\gamma}{1-\alpha_1\gamma}\alpha_2.
\end{align*}
Define $\bar x_k = (1/(k+1))\sum_{l=0}^k x_l$. Again by the convexity of $f$, we have
\begin{align*}
\mathbf{E} \{  f(\bar x_k) - f(x^\star)\} 
& \leq \frac{1}{k+1} \sum_{l=0}^k\mathbf{E} \{  f( x_l) - f(x^\star)\} \\  
&\leq \frac{1}{k+1}\frac{1}{\gamma} \frac{1}{1-\alpha_1\gamma}\left( \mathbf{E}\| \tilde x_0  -x^\star \|^2 - \mathbf{E}\| \tilde x_{k+1}  -x^\star \|^2 \right) \\
& \hspace{0.4cm}+ \frac{2L}{1-\alpha_1\gamma}\beta + \frac{\gamma}{1-\alpha_1\gamma}\alpha_2.
\end{align*}
Since $\tilde x_0 = x_0$ and $\|x\|^2\geq 0 $ for $x\in\mathbb{R}^d$, we complete the proof. 
\end{proof}


\begin{lemma}\label{lemma:NonConvex_EC}
Consider the optimization problem \eqref{eqn:Problem} where $f$ is $L-$smooth. Assume that $g_k:\mathbb{R}^d\rightarrow\mathbb{R}^d$ satisfies  
\begin{align*}
\mathbf{E}\{g_k \}= \nabla f(x_k)  \quad \text { and  } \quad \mathbf{E} \| g_k\|^2 \leq \alpha_1 \mathbf{E}\| \nabla f(x_k) \|^2 + \alpha_2,
\end{align*}
for positive constants $\alpha_1,\alpha_2$. Then, the iterates $\{x_k\}_{k\in\mathbb{N}}$ generated by the recursion \[\tilde x_{k+1} = \tilde x_k - \gamma g_k\] 
where $\tilde x_0 = x_0$, $\mathbf{E}\| x_k -\tilde x_k\|^2\leq\beta$ for a positive constant $\beta$, and  $\gamma <1/(\alpha_1 L)$ satisfy
\begin{align*}
\min_{l\in\{0,1,\ldots,k\}}\mathbf{E} \| \nabla f(x_l) \|^2 
\leq \frac{1}{k+1}\cdot \frac{2}{\gamma} \frac{1}{1-\alpha_1L\gamma} \left( \mathbf{E} f(\tilde x_0) - \mathbf{E} f(\tilde x_{k+1})  \right) + \frac{L}{1-\alpha_1L\gamma}(L\beta+\gamma\alpha_2).
\end{align*}
\end{lemma}
\begin{proof}
By the smoothness assumption of $f$ and by the recursion, 
\begin{align*}
f(\tilde x_{k+1}) \leq f(\tilde x_k) - \gamma \left\langle \nabla f(\tilde x_k),  g_k  \right\rangle + \frac{L\gamma^2}{2} \left\| g_k \right\|^2.
\end{align*}
Taking the expectation over the randomness in the algorithm yields
\begin{align*}
\mathbf{E} f(\tilde x_{k+1}) 
&\leq \mathbf{E} f(\tilde x_k) - \gamma \mathbf{E}\left\langle \nabla f(\tilde x_k),  \nabla f(x_k)  \right\rangle + \frac{L\gamma^2}{2} \mathbf{E}\left\| g_k \right\|^2.
\end{align*}
By the fact that $-2\langle x,y \rangle = -\| x \|^2 -\| y\|^2+ \| x-y\|^2$ for $x,y\in\mathbb{R}^d$ and by the second property of $g_k$, we have: 
\begin{align*}
\mathbf{E} f(\tilde x_{k+1}) 
&\leq \mathbf{E} f(\tilde x_k) - \frac{\gamma}{2} \mathbf{E}\| \nabla f(\tilde x_k) \|^2 -\left( \frac{\gamma}{2}  - \frac{\alpha_1 L\gamma^2}{2}  \right)\mathbf{E}\| \nabla f( x_k) \|^2 + \frac{\gamma}{2} \mathbf{E}\| \nabla f( x_k) -\nabla f(\tilde x_k) \|^2 \\
&\hspace{0.4cm}+ \frac{L\gamma^2}{2}\alpha_2.
\end{align*}
Assume that $\mathbf{E}\| x_k - \tilde x_k \|^2 \leq \beta$ for the positive constant $\beta$ and that $\gamma<1/(\alpha_1 L)$. By the smooothness of $f$, we have
\begin{align*}
\mathbf{E} f(\tilde x_{k+1}) 
&\leq \mathbf{E} f(\tilde x_k)  -\left( \frac{\gamma}{2}  - \frac{\alpha_1 L\gamma^2}{2}  \right)\mathbf{E}\| \nabla f( x_k) \|^2 + \frac{L^2\gamma}{2}\beta + \frac{L\gamma^2}{2}\alpha_2,
\end{align*}
or equivalently 
\begin{align*}
\mathbf{E}\| \nabla f( x_k) \|^2 \leq \frac{2}{\gamma} \frac{1}{1-\alpha_1L\gamma} \left( \mathbf{E} f(\tilde x_k) - \mathbf{E} f(\tilde x_{k+1})  \right) + \frac{L}{1-\alpha_1L\gamma}(L\beta+\gamma\alpha_2).
\end{align*}
Therefore, 
\begin{align*}
\min_{l\in\{0,1,\ldots,k\}}\mathbf{E} \| \nabla f(x_l) \|^2 
& \leq \frac{1}{k+1} \sum_{l=0}^k \mathbf{E} \| \nabla f(x_l) \|^2 \\
&\leq \frac{1}{k+1}\cdot \frac{2}{\gamma} \frac{1}{1-\alpha_1L\gamma} \left( \mathbf{E} f(\tilde x_0) - \mathbf{E} f(\tilde x_{k+1})  \right) + \frac{L}{1-\alpha_1L\gamma}(L\beta+\gamma\alpha_2).
\end{align*}
Since $\tilde x_0 = x_0$ and $f(x)\geq f(x^\star)$ for $x\in\mathbb{R}^d$, the proof is complete.
\end{proof}